\documentclass[twocolumn,superscriptaddress,amssymb,amsmath,
nobibnotes,aps,prd,showpacs,nofootinbib]{revtex4}%
\usepackage{graphicx}
\usepackage{epsf}
\usepackage{bm}
\usepackage{amsmath}
\usepackage{amsfonts}
\usepackage{amssymb}
\usepackage{epstopdf}
\usepackage{natbib}%
\usepackage{rotating}
\usepackage{pdflscape}
\providecommand{\U}[1]{\protect\rule{.1in}{.1in}}
\newcommand{\be}{\begin{equation}}
\newcommand{\ee}{\end{equation}}

\newcommand{\mincir}{\raise
-3.truept\hbox{\rlap{\hbox{$\sim$}}\raise4.truept\hbox{$<$}\ }}
\newcommand{\magcir}{\raise
-3.truept\hbox{\rlap{\hbox{$\sim$}}\raise4.truept\hbox{$>$}\ }}

\usepackage{color}

\begin{document}

\title{Effects of Anisotropic Stress in Interacting Dark Matter - Dark Energy Scenarios}

\author{Weiqiang Yang}
\email{d11102004@163.com}
\affiliation{Department of Physics, Liaoning Normal University, Dalian, 116029, P. R.
China}
\author{Supriya Pan}
\email{span@research.jdvu.ac.in}
\affiliation{Department of Mathematics, Raiganj Surendranath Mahavidyalaya, Sudarshanpur, Raiganj, Uttar Dinajpur, West Bengal 733134, India}
\author{Lixin Xu}
\email{lxxu@dlut.edu.cn}
\affiliation{Institute of Theoretical Physics, School  of Physics, Dalian  University  of  Technology,  Dalian,  116024, P.  R.  China}
\author{David F. Mota}
\email{mota@astro.uio.no}
\affiliation{Institute of Theoretical Astrophysics, University of Oslo, P.O. Box 1029 Blindern, N-0315 Oslo, Norway}

\begin{abstract}
We study a novel interacting dark energy $-$ dark matter scenario where the anisotropic stress of the large scale inhomogeneities is considered. The dark energy has a constant equation of state and the interaction model produces stable perturbations.
The resulting picture is constrained using different astronomical data aiming to measure the impact of the anisotropic stress on the cosmological parameters. Our analyses show that a non-zero interaction in the dark sector is allowed while a non-interaction scenario is recovered within 68\% CL. The anisotropic stress is also constrained to be small, and its zero value is permitted within 68\% CL. The dark energy equation of state, $w_x$, is also found to be close to `$-1$' boundary. However, from the ratio of the CMB TT spectra, we see that the model has a mild deviation from the $\Lambda$CDM cosmology while such deviation is almost forbidden from the CMB TT spectra alone. Although the deviation is not much significant, but from the present data, we cannot exclude such deviation. Overall, at the background level, the model is close to the $\Lambda$CDM cosmology while at the level of perturbations, a non-zero but a very small interaction in the dark sector is permitted. Perhaps, a more accurate conclusion can be made with the next generation of surveys.  We also found that the region $w_x < -1$, is found to be effective to release the tension on $H_0$. Finally, from the Bayesian analysis, we find that $\Lambda$CDM remains in still preferred over the interacting scenarios.   
\end{abstract}

\maketitle

\section{Introduction}

A remarkable revolution in the dynamical history of the universe has been witnessed in the last several years. Around 20 years back, the distant Supernovae of Type Ia (SNIa) first indicated an accelerating expansion of the universe and thereafter a lot of distinct  astronomical observations have strengthened such observational prediction. To interpret this acceleration a hypothetical fluid with negative pressure became necessary and subsequently, cosmological constant was revived into the picture. The cosmological constant, $\Lambda$, has a negative equation of state, $P_{\Lambda} = -\rho_{\Lambda}$ and together with cold-dark-matter, the joint scenario $\Lambda$CDM has been found to be the best cosmological model, at least according to a series of astronomical measurements. In $\Lambda$CDM scenario, both the cosmological constant and the cold-dark-matter remain conserved separately, as if they are two disjoint sectors. Such model of the universe is widely referred to as the non-interacting cosmological scenario. Unfortunately, the cosmological constant suffers from the fine-tuning problem (also known as the cosmological constant problem), where being time-independent, it reports an unimaginable difference (of the order of $10^{121}$) in its value determined in the Planck and low energy scales. The problem associated with the cosmological constant is not a new detection, it is persisting since long \cite{Weinberg1989}, even before the late-accelerating phase.  Thus, attempts have been made aiming to provide with a reasonable justification on the fine-tuning problem \cite{Wetterich:1994bg}. While  on the other hand, people have tried to bypass this problem through the introduction of dark energy models \cite{Copeland:2006wr, AT, Bamba:2012cp} (we also refer to  
some specific scalar field dark energy models that may also account for the early scenarios of the universe \cite{deHaro:2016hpl, deHaro:2016cdm}). 
However, although the introduction of dark energy models relieves the cosmological constant problem but they raised another serious issue which is widely known as the cosmic coincidence problem \cite{Zlatev:1998tr}. Such coincidence problem led to another class of cosmological theories which is the theory of non-gravitational interaction between dark matter and dark energy.

The non-gravitational interaction in the dark sector, precisely between dark matter and dark energy is a phenomenological concept that was originally thought to explain the different values of the time-independent cosmological constant \cite{Wetterich:1994bg}, but later on, such concept was found to be very useful to explain the cosmic coincidence problem \cite{Amendola:1999er, Chimento:2003iea,Cai:2004dk, Hu:2006ar, delCampo:2008sr,delCampo:2008jx}. Certainly, this led to a large amount of investigations towards this direction where the dark sectors have direct interaction \cite{Billyard:2000bh,Barrow:2006hia, Chimento:2009hj,Quartin:2008px,Valiviita:2009nu,Thorsrud:2012mu, dfm1,dfm2,dfm3,dfm4,dfm5,Yang:2014hea, Yang:2014gza, yang:2014vza, Faraoni:2014vra, Duniya:2015nva, Valiviita:2015dfa, Pan:2012ki, Pan:2016ngu, Mukherjee:2016shl, Sharov:2017iue, Santos:2017bqm, Pan:2017ent} (also see \cite{Pan:2013rha, Chen:2011cy, Pan:2014afa, Shahalam:2015sja, Shahalam:2017fqt, Cai:2017yww, Kumar:2017bpv, Odintsov:2017icc}). Such interacting scenarios have good motivation if the particle physics theory is considered, because from the particle physics view, mutual interaction between any two fields, is a natural phenomenon, irrespective of the nature of the fields. Although the interacting dynamics is complicated and a generalized cosmic scenario, but it recovers the non-interaction cosmology as a special case. Thus, the theory of non-gravitational interaction between dark matter and dark energy is a generalized version of the non-interacting dark matter and dark energy cosmologies.  Interestingly enough, the observational data at recent time found that the direct interaction between dark matter and dark energy cannot be excluded \cite{Salvatelli:2014zta, Nunes:2016dlj, Yang:2016evp, Kumar:2016zpg, vandeBruck:2016hpz, Yang:2017yme, Kumar:2017dnp, Yang:2017zjs, Yang:2017ccc}. Moreover, very recently, it has been reported that the current tension on the local Hubble constant can be alleviated with the introduction of dark matter and dark energy interactions \cite{DiValentino:2017iww, Kumar:2016zpg}. Additionally, the crossing of phantom barrier has also been found to be an easy consequence of the non-gravitational interaction. Thus, the theory of interacting dark energy might be considered to be an appealing field of research and indeed a hot topic for the next generation of the astronomical surveys.

This work presents a general interacting scenario where besides from the non-gravitational interaction between dark matter and dark energy, we also include the possibility of an anisotropic stress. The anisotropic stress appears when the first
order perturbation is considered, and in most of the cases, it is generally neglected.
From both theoretical and the observational grounds, the possibility of an anisotropic stress cannot be excluded at all. Although the dimension of the 
resulting parameters space is increased, but, due to advancements of the astronomical data, the measurement of the anisotropic stress becomes important. The most important thing is to measure the effect of this quantity on the large scale structure evolution of the concerned cosmological scenario. This is the primary motivation of this work.  In particular, looking at the perturbation equations (see section \ref{sec-bg+per}) one can realize that a nonzero value of the anisotropic stress, can affect the  temperature anisotropy in the cosmic microwave background spectra and also on the matter power spectra as well. Thus, for a detailed understanding of the interacting scenario in its large scale structure, the anisotropic stress plays a significant role.

Thus, following the above motivation, in a spatially flat Friedmann-Lema\^itre-Robertson-Walker universe, we study an
interacting dark matter-dark energy scenario in presence of an anisotropic stress and then constrain this model using a series of latest astronomical data from \textit{cosmic microwave background radiation}, \textit{Joint Light Curve analysis} (JLA) sample of Supernovae Type Ia, \textit{baryon acoustic oscillations} (BAO) distance measurements, \textit{Hubble parameter measurements} from cosmic chronometers (CC), \textit{weak gravitational lensing} and the \textit{local Hubble constant value} from the Hubble Space Telescope (HST). The analyses are based on the use of publicly available markov chain monte carlo package \texttt{cosmomc} where the convergence of the cosmological parameters follows the well known Gelman-Rubin statistics.

The work has been organized in the following way.   In section \ref{sec-bg+per} we describe the background and the perturbation equations for the coupled dark energy in presence of the matter-sourced anisotropic stress. Section \ref{sec-data} describes the observational data that we use to analyze the present models. In section \ref{sec-results} we discuss the constraints on the current model. Finally, section \ref{sec-discuss} closes the work with the main findings of this investigation.

\section{The interacting universe}
\label{sec-bg+per}

In this section we shall describe an interacting cosmological scenario both at background and perturbative levels. To do this, we assume the most general metric for the underlying geometry of the universe which is characterized by the  Friedmann-Lema\^{i}tre-Robertson-Walker (FLRW) line element. We also assume that the metric is spatially flat. The evolution equations for a pressureless dark matter and a dark energy fluid in this universe obey the following conservation equations

\begin{eqnarray}
\rho'_c+3\mathcal{H}\rho_c =  a Q_c = -aQ, \label{cons1}\\
\rho'_x+3\mathcal{H}(1+w_x)\rho_x = a Q_x =aQ, \label{cons2}
\end{eqnarray}
where the prime denotes the differentiation with respect to the conformal time $\tau$ (i.e. ``$~ ' \equiv \frac{d}{d\tau}$''); $\mathcal{H}=a'/a$, is the conformal Hubble parameter; $\rho_c$, $\rho_x$ are respectively the energy densities of cold dark matter and dark energy; $w_x$ is the barotropic equation of state of the dark energy, that means $w_x= p_x/\rho_x$, here $p_x$ is the pressure of the dark energy fluid and as usual zero pressure is attributed to cold-dark-matter sector. The quantity $Q \, (= Q_x=-Q_c)$ in the right hand sides of (\ref{cons1}) and (\ref{cons2}) is the energy transfer rate between the dark sectors and depending on its sign the direction of energy flow is determined. To be precise, a positive interaction rate ($Q >0$) assigns the energy flow from dark matter to dark energy while the negative interaction rate reverses its direction of energy flow.
In addition, we consider the presence of non-relativistic baryons ($\rho_b$) and relativistic radiation ($\rho_r$) which follow the standard evolution equations, that means they do not take part in the interaction. The dynamics of the spatially flat universe is thus constrained by the Hubble equation
\begin{eqnarray}\label{Friedmann}
\mathcal{H}^2 = \frac{8 \pi G}{3} a^2 \rho_{tot},
\end{eqnarray}
where $\rho_{tot}=\rho_c+\rho_x+\rho_b+\rho_r$, is the total energy density of the universe. Hence, if the energy transfer rate, $Q$, is specified, then the evolution equations for $\rho_c$ and $\rho_x$ can fully be determined using the conservation equations (\ref{cons1}) and (\ref{cons2}) together with the Hubble constraint (\ref{Friedmann}).
However, the presence of interaction in the dark sector may
significantly affect the large scale structure of the universe and hence it becomes necessary to consider the evolution equations for the interacting model at the perturbative levels. Thus, in order to do so we consider the perturbed FLRW metric \cite{Mukhanov, Ma:1995ey, Malik:2008im}

\begin{eqnarray}
ds^2=a^2(\tau) \Bigg[ -(1+2\phi)d\tau^2 + 2\partial_i B d\tau dx^i \nonumber\\+ \Bigl((1-2\psi)\delta_{ij}+2\partial_i\partial_jE \Bigr)dx^idx^j \Bigg].
\label{eq:per-metric}
\end{eqnarray}
Here, the quantities appearing in the above metric (\ref{eq:per-metric}), namely, $\phi$, $B$, $\psi$ and $E$ represent the gauge-dependent scalar perturbations. For this perturbed metric (\ref{eq:per-metric}), using the consdervation equations $\nabla _{\nu }T_{A}^{\mu \nu }=Q_{A}^{\mu },$ where $\sum\limits_{\mathrm{A}}{%
Q_{A}^{\mu }}=0$, one can derive the perturbation equations for the dark fluids characterized by the symbol $A$ (for cold dark matter $A= c$, and for dark energy $A = x$). Here, $Q_{A}^{\mu }=(Q_{A}+\delta Q_{A})u^{\mu }+F_A^{\mu}$, where 
$Q_A$ presents the transfer rate of the energy flow between the dark fluids 
and $F_A^{\mu} = a^{-1} (0, \partial^{i}f_A)$, is the momentum density transfer relative to the four-velocity vector $u^{\mu}$ in which $f_A$ is the momentum transfer potential.  Now, following \cite{Valiviita:2008iv, Majerotto:2009np}, the perturbed energy and momentum balance equations for the interacting dark matter and dark energy scenario one can write 
\begin{eqnarray}
\delta \rho'_A+3\mathcal{H}(\delta \rho_A+\delta p_A)-3(\rho_A+p_A)\psi'\nonumber\\-k^2(\rho_A+p_A)(v_A+E')
=aQ_A\phi+a\delta Q_A,
\end{eqnarray}
\begin{eqnarray}
\delta p_A+[(\rho_A+p_A)(v_A+B)]'+4\mathcal{H}(\rho_A+p_A)(v_A+B) \nonumber\\+(\rho_A+p_A)\phi-\frac{2}{3}k^2p_A \Pi_A
=aQ_A(v + B)+af_A,
\label{eq:delta-PA}
\end{eqnarray}
where prime stands for the 
differentiation with respect to the conformal time, mentioned earlier; 
$\mathcal{H}$ is the conformal Hubble rate; the quantity $\Pi_A$ is related to the anisotropic stress $\sigma_A$ of the fluid $A$. 
The relation between the peculiar velocity potential $v_A$ and the local volume expansion rate $\theta_A$ is, $\theta_A=-k^2(v_A+B)$ in the Fourier space with mode $k$ \cite{Mukhanov, Ma:1995ey, Malik:2008im, Valiviita:2008iv, Kodama:1985bj}. Here by $\delta_A=\delta\rho_A/\rho_A$, we mean the density perturbation for the fluid $A$, and momentum transfer potential $f_A$ has been assumed to be the simplest physical choice, that gives its value to zero in the rest frame of dark matter
\cite{Valiviita:2008iv,Koyama:2009gd,Clemson:2011an}. Hence, the momentum transfer potential takes the expression $k^2f_A=Q_A(\theta-\theta_c)$, see \cite{Yang:2014gza}, where $\theta = \theta_{\mu}^{\mu}$, is the volume expansion of the total fluid and $\theta_c$ is the volume expansion of the cold dark matter fluid.

Now, the pressure perturbation $\delta p_A$, for any fluid $A$, is related as $\delta p_A=c^2_{sA}\delta\rho_A+(c^2_{sA}-c^2_{aA})\rho'_A(v_A+B)$ \cite{Valiviita:2008iv} which for the dark energy fluid turns out to be $\delta p_x=c^2_{sx}\delta\rho_x+(c^2_{sx}-c^2_{ax})\left[3\mathcal{H}(1+w_x)\rho_x-aQ\right]\theta_x/k^2$.

The evolution of fluid perturbations could be described by the adiabatic speed of sound $c^2_{aA}\equiv p'_A/\rho'_A=w_A-w'_A/[3\mathcal{H}(1+w_A)]$. In this adiabatic case, the relation between the perturbations of $\delta p_A$ and $\delta\rho_A$ is related by $\delta p_A=c^2_{aA}\delta\rho_A$. However, for an entropic fluid, the pressure might not be a unique function of the energy density $\rho_A$. Therefore, there would be another degree of freedom to describe the microproperties of a general fluid. That is the physical speed of sound in the rest frame $c^2_{sA}\equiv(\delta p_A/\delta\rho_A)|_{rf}$ (`${rf}$' represents the rest-frame) which is defined in the comoving frame of the fluid. Now, we note that when the entropic perturbation vanishes, the physical sound speed and the adiabatic sound speed vanishes, that means, $c^2_{sA}=c^2_{aA}$. Hence, in the case of entropic fluid such as scalar fields, one needs both its equation of state and its sound speed, to have a complete description of dark energy and its perturbations. However, in order to
fully describe a dark energy fluid and its perturbations, one should also consider the possibility of an anisotropic stress, even in an isotropic and homogeneous FLRW universe, where the anisotropic stress $\sigma_A = \frac{2w_A}{3(1+w_A)}\Pi_A$, can be taken as a spatial perturbation. 

In the synchronous gauge ($\phi= B = 0$, $\psi= \eta$, and $k^2E = -h/2-3\eta$), the evolution equations for density perturbations and velocity perturbations equations for dark energy and dark matter respectively read
\begin{eqnarray}
\delta'_x
=-(1+w_x)\left(\theta_x+\frac{h'}{2}\right) -3\mathcal{H}w'_x\frac{\theta_x}{k^2} \nonumber\\
-3\mathcal{H}(c^2_{sx}-w_x)\left[\delta_x+3\mathcal{H}(1+w_x)\frac{\theta_x}{k^2}\right]
\nonumber \\
+\frac{aQ}{\rho_x}\left[-\delta_x+\frac{\delta Q}{Q}+3\mathcal{H}(c^2_{sx}-w_x)\frac{\theta_x}{k^2}\right], \label{per1}
\end{eqnarray}
\begin{eqnarray}
\theta'_x
=-\mathcal{H}(1-3c^2_{sx})\theta_x+\frac{c^2_{sx}}{(1+w_x)}k^2\delta_x
-k^2\sigma_x \nonumber\\+\frac{aQ}{\rho_x}\left[\frac{\theta_c-(1+c^2_{sx})\theta_x}{1+w_x}\right], \label{per2}
\end{eqnarray}
\begin{eqnarray}
\delta'_c
&=&-\left(\theta_c+\frac{h'}{2}\right)
+\frac{aQ}{\rho_c}\left(\delta_c-\frac{\delta Q}{Q}\right), \label{per3}\\
\theta'_c
&=&-\mathcal{H}\theta_c, \label{per4}
\label{eq:perturbation}
\end{eqnarray}
where the factor $\delta Q/Q$ includes the perturbation term of the Hubble expansion rate $\delta H$ \cite{Gavela:2010tm}.  The effects of anisotropic stress (present in equation (\ref{per2})) can be studied in two distinct ways as follows.  The first approach is to assume a parametrized differential equation for $\sigma_x$ given by \cite{Hu:1998kj}:

\begin{equation}
\sigma'_x+3\mathcal{H}\frac{c^2_{ax}}{w_x}\sigma_x
=\frac{8}{3}\frac{c^2_{vis}}{1+w_x}\left(\theta_x+\frac{h'}{2}+3\eta'\right),
\label{eq:shear}
\end{equation}
where $c^2_{vis}$ is the viscous speed of sound which controls the correspondence between the velocity (or metric) shear and the anisotropic stress. In particular, for a relativistic fluid, $c^2_{vis} = 1/3$, while for a general dark energy fluid, $c^2_{vis}$ is a free parameter and it can be constrained through the observational data \cite{Huey:2004jz, Koivisto:2005mm, Mota:2007sz,Koivisto:2008ig, Cardona:2014iba, Amendola:2013qna, Saltas:2010tt, Kunz:2006ca,%
Song:2010rm}.  Secondly, one may directly assume an appropriate expression for 
$\sigma_x$ since the anisotropic stress  (external or internal) can be linked to the overdensity of  matter (consequently dark matter) or dark energy as shown in \cite{Cardona:2014iba}. In this work we shall follow the second approach, that means, the anisotropic stress linked to the overdensity of dark matter/dark energy.
 Now, concerning the interacting cosmologies, the presence of externally sourced anisotropic stress or the internally sourced anisotropic stress are equally favored and none of the possibilities have been studied so far. So, for the first time, we begin this new analysis with the externally sourced anisotropic stress leaving the second possibility as a future work in this direction. The external anisotropic stress that is linked to the overdensity of dark matter is also known as the matter-sourced anisotropic stress model having the form
\begin{eqnarray}
\sigma_x=\frac{2}{3}\frac{1}{1+w_x}e_{\pi}a^n\Delta_m,
\label{eq:matter-sourced}
\end{eqnarray}
while when the anisotropic stress is linked to the overdensity of dark energy (similarly, it might be dubbed as the dark-energy-sourced anisotropic stress model)
\begin{eqnarray}
\sigma_x=\frac{2}{3}\frac{1}{1+w_x}\frac{f_{\pi}}{1+(g_{\pi}\mathcal{H}/k)^2}\Delta_x,
\label{eq:dark-energy-sourced}
\end{eqnarray}
where $\Delta_i=\delta_i-\frac{\rho'_i}{\rho_i}\frac{\theta_i}{k^2}$, is the gauge invariant density perturbations for matter ($i=m$) and dark energy ($i=x$), respectively. The above relations are established on the fact that the anisotropic stress and the overdensity of dark matter (or, dark energy) may modify the gravitational slip in an effective way  \cite{Song:2010rm}. The latest analysis on the observational constraints of dark energy with anisotropic stress can be found in \cite{Chang:2014mta, Chang:2014bea}.

Let us come to the interaction model that we wish to study in this work. Before taking any typical interaction model we recall that the interaction function $Q$ directly enters into the pressure perturbation for dark energy as \cite{Valiviita:2008iv}
\begin{align}
\delta p_x
=c^2_{sx}\delta\rho_x+(c^2_{sx}-c^2_{ax})\left[3\mathcal{H}(1+w_x)\rho_x-aQ\right]\frac{\theta_x}{k^2}
\label{eq:deltap}
\end{align}

According to the qualitative analysis on the large-scale instability in the dark sector perturbations during the early radiation era \cite{Valiviita:2008iv}, in the pressure perturbation of dark energy (\ref{eq:deltap}), the coupling term $Q$ in the pressure perturbation $\delta p_x$ can lead to a driving term $\frac{aQ}{\rho_x}\left[\frac{\theta_c-(1+c^2_{sx})\theta_x}{1+w_x}\right]$ which includes the factor $\mathcal{H}\theta_x$, and it becomes very large if $w_x$ is close to `$-1$'. This causes rapid growth of $\theta_x$. Qualitatively, this is the source of the instability: in the presence of energy-momentum transfer in the perturbed dark fluids, momentum balance requires a runaway growth of the dark energy velocity. In order to avoid the perturbation instability, and based on the phenomenological consideration, we assume the constant equation of state $w_x$ in the interacting dark energy with the energy transfer rate $Q=3H\xi(1+w_x)\rho_x$. The presence of the factor $(1+w_x)$ in the interaction function does not bother with the dark energy equation of state, and hence the stability of the interaction model in the large scale structure of the universe rests on the coupling parameter of the interaction. The perturbation equations (\ref{per1}) to (\ref{per4}) for the specific interaction model turn out to be
\begin{eqnarray}
\delta'_x
=-(1+w_x)\left(\theta_x+\frac{h'}{2}\right)
\nonumber\\-3\mathcal{H}(c^2_{sx}-w_x)\left[\delta_x+3\mathcal{H}(1+w_x)\frac{\theta_x}{k^2}\right] \nonumber \\
+3\mathcal{H}\xi(1+w_x)\left[\frac{\theta+h'/2}{3\mathcal{H}}+3\mathcal{H}(c^2_{sx}-w_x)\frac{\theta_x}{k^2}\right],\nonumber
\end{eqnarray}
\begin{eqnarray}
\theta'_x
=-\mathcal{H}(1-3c^2_{sx})\theta_x+\frac{c^2_{sx}}{(1+w_x)}k^2\delta_x \nonumber\\-k^2\sigma_x
+3\mathcal{H}\xi\left[\theta_c-(1+c^2_{sx})\theta_x\right],\nonumber
\end{eqnarray}
\begin{eqnarray}
\delta'_c
&=&-\left(\theta_c+\frac{h'}{2}\right)
+3\mathcal{H}\xi(1+w_x)\frac{\rho_x}{\rho_c}\left(\delta_c-\delta_x-\frac{\theta+h'/2}{3\mathcal{H}}\right), \nonumber\\
\theta'_c
&=&-\mathcal{H}\theta_c,\nonumber
\label{eq:perturbation1}
\end{eqnarray}
where the matter-sourced anisotropic stress is, $\sigma_x=2/[3(1+w_x)]e_{\pi}a^n\Delta_m$. In this work, we consider the matter-sourced model with $n=0$, that means,
$\sigma_x=2/[3(1+w_x)]e_{\pi}\Delta_m$, as the simplest case in such complicated interacting dynamics. Although there is no such strict restriction to exclude the possibility of anisotropic stress sourced by dark energy, but, however, since the cluster effects of dark energy is smaller in compared to the dark matter, the effects of anisotropic stress sourced by dark energy must be weaker in respect to the anisotropic stress sourced by dark matter. As a result, the anisotropic stress sourced by dark matter might be more relevant in this context.

\section{Observational data sets and the statistical technique}
\label{sec-data}

In this section we describe the main observational data that we have used to constrain the cosmological
scenarios and also we outline the statistical methodology.
We use various astronomical data ranging from low redshifts to high redshifts, for our analysis. Below we summarize the data sets with their corresponding references.

\begin{enumerate}

\item \textit{Cosmic Microwave Background Radiation:} The full Planck 2015 low$-l$ temperature-plus-polarization and the high$-l$ $C^{TE}_l+C^{EE}_l$ likelihood (``Planck TT, TE, EE + lowTEB'') \cite{Adam:2015rua, Aghanim:2015xee} have been used. For the interacting dark energy with matter-sourced anisotropic stress, the amplitude of CMB at low multipole ($l<30$) is very sensitive to the values due to the fact that the anisotropic stress of dark energy is proportional to the overdensity of dark matter directly. The summation of the Newtonian potentials becomes
    \begin{eqnarray}
    k^2(\Phi+\Psi)=-8\pi Ga^2\left(\sum\limits_A\rho_A\Delta_A+\sum\limits_Ap_A \Pi_A\right),
    \label{eq:potential}
    \end{eqnarray}
    where $\Delta_A$ is the gauge invariant density contrast and $\Pi_A$ is related to the anisotropic stress $\sigma_A$ via $\sigma_A=\frac{2}{3}\frac{w_A}{1+w_A}\Pi_A$. Thus, an extra contribution to the integrated Sachs$-$Wolfe (ISW) effect due to the existence of anisotropic stress of dark energy is
    \begin{eqnarray}
    -k^2ISW_{stress}=8\pi Ga^2\sum\limits_A p_A \dot{\Pi}_A
    -8\pi Ga^2\mathcal{H}\Bigg[4\sum\limits_Ap_A \Pi_A \nonumber\\ + \sum\limits_A(3\rho_A-p_A) \Pi_A -\sum\limits_A \frac{d\ln w_A}{d\ln a}p_A\Pi_A \Bigg].
    \nonumber\\
    \label{eq:ISW}
    \end{eqnarray}

\item \textit{Joint Light-Curve Analaysis:} The Joint Light-curve Analysis (JLA) sample \cite{Betoule:2014frx} containing 740 Supernovae Type Ia in the low-redshift range $z\in[0.01, 1.30]$ have been considered.

\item \textit{Baryon Acoustic Oscillations Distance Measurements:} For baryon acoustic oscillations (BAO) data, we mainly use four different data points. In particular, we use the CMASS and LOWZ samples from the latest Data Release 12 (DR12) of the Baryon Oscillation Spectroscopic Survey (BOSS) respectively at the effective redshifts $z_{\rm eff}=0.57$ and $z_{\rm eff}=0.32$~\cite{Gil-Marin:2015nqa}. In addition, we include the 6dF Galaxy Survey (6dFGS) measurement at $z_{\emph{\emph{eff}}}=0.106$~\cite{Beutler:2011hx}, and the Main Galaxy Sample of Data Release 7 of Sloan Digital Sky Survey (SDSS-MGS) at $z_{\emph{\emph{eff}}}=0.15$~\cite{Ross:2014qpa}.

\item \textit{Redshift Space Distortion Data:} We employ the redshift space distortion (RSD) measurements from two disctinct galaxy samples, the one which includes the CMASS sample with an effective redshift of $z_{\rm eff}=0.57$ \cite{Gil-Marin:2016wya} while the other includes the LOWZ sample with an effective redshift of $z_{\rm eff}=0.32$~\cite{Gil-Marin:2016wya}.

\item \textit{Hubble Parameter Measurements:} We also employ the recently released cosmic chronometers (CC) data with $30$ measurements of the Hubble parameter values in the redshift interval $0 < z< 2$ \cite{Moresco:2016mzx}. The cosmic chronometers are basically some galaxies which evolve passively and  are the most massive. An accurate measurement of the differential age evolutions $dt$ of such galaxies together with the spectroscopic estimation of $dz$ with high accuracy yield the Hubble parameter value through $H (z) = (1+z)^{-1} dz/dt$. For more on the CC, we refer the readers to \cite{Moresco:2016mzx}.

\item \textit{$H_0$ from the Hubble Space Telescope}: The present Hubble constant value yielding $H_0=73.02\pm1.79 km s^{-1} Mpc^{-1}$ \cite{Riess:2016jrr}  from the Hubble Space Telescope (HST) has been used. We label this value as HST.

\item \textit{Weak Gravitational Lensing Data:} Finally, we also use the weak gravitational lensing data (WL) along with the previous data sets. The sample is taken from the Canada-France-Hawaii Telescope Lensing Survey (CFHTLenS) which spans 154 square degrees in five optical bands. In this survey, 21 sets of cosmic shear correlation functions linked to six redshift bins have been presented, see  Refs. \cite{Heymans:2013fya, Asgari:2016xuw} for details. The tomographic correlation functions measured  measured from the blue galaxy sample and consistent with zero intrinsic alignment nuisance parameter has been named as $blue\_sample$ and we have used this $blue\_sample$ for the present work. From the likelihood analysis of the CFHTLenS data one can extract the information of our Universe. Here, the true inverse covariance matrix takes the form $\mathbb{C}^{-1}=\alpha_A\hat{\mathbb{C}}^{-1}$ in which $\alpha_A=(n_{\mu}-p-2)/(n_{\mu}-1)$, and $\hat{\mathbb{C}}$ is the measured covariance matrix. The inclusion of the anisotropic stress of dark energy certainly modifies the summation of potentials given in Eq. (\ref{eq:potential}). Moreover, for the presence of anisotropic stress, the lensing potential gains an extra contribution leading to the convergence power spectrum
at angular wave number $l$ as 
\begin{eqnarray}
    P^{ij}_K(l)=\int^{\eta_H}_0 d\eta\frac{q_i(\eta)q_j(\eta)}{[f_K(\chi)]^2}
    \left(1+\frac{\sum_Ap_A\Pi_A}{\sum_A\rho_A\Delta_A}\right)^2P_{\delta}
    \left(k=\frac{l}{f_K(\eta)};\eta\right),
    \label{eq:Pij}
    \end{eqnarray}
    where $\eta$ is the comoving distance; $f_K(\eta)$ is the angular diameter distance out to $\eta$ and it depends on the curvature scalar $K$. We note that in the present work we have assumed $K= 0$. The quantity $\eta_H$ is the horizon distance, and $q_i(\eta)$ represents lensing efficiency function for the redshift bin $i$, see \cite{Heymans:2013fya, Asgari:2016xuw} for more discussions.
    
\end{enumerate}

Now, for the interacting dark energy with matter-sourced anisotropic stress, the amplitude of CMB at low multipole ($l<30$) is very sensitive to the values of $e_{\pi}$ due to the fact that the anisotropic stress of dark energy is proportional directly to the overdensity of dark matter. The summation of the Newtonian potentials becomes
\begin{eqnarray}
k^2(\Phi+\Psi)
=-8\pi Ga^2\sum\limits_A\rho_A\Delta_A-8\pi Ga^2\sum\limits_Ap_A\Pi_A \nonumber \\
=-8\pi Ga^2(\rho_b\Delta_b+\rho_c\Delta_c+\rho_x\Delta_x+p_x\pi_x) \nonumber \\
=-8\pi Ga^2 \Bigg[\rho_b\delta_b+\rho_c\delta_c+\rho_x\delta_x
+\Bigg(3\mathcal{H}(1+w_x)\rho_x \nonumber\\-\frac{1}{2}(5+3c^2_{sx})aQ \Bigg)\frac{\theta_x}{k^2} \Bigg]
\label{eq:potentialde}
\end{eqnarray}
where $\pi_x$ is related to the anisotropic stress $\sigma_x$ through the relation $\sigma_x=\frac{2w_x}{3(1+w_x)}\pi_x$.
For the influence of WL, the convergence power spectrum will also be modified by the anisotropic stress in the same way, but in the spatial part of the Newtonian potentials.

The likelihood for our analysis is, $\mathcal{L}\propto e^{-\chi^2_{tot}/2}$, where $\chi^2_{tot}$ is,
$\chi^2_{tot}=\chi^2_{JLA}+\chi^2_{BAO}+\chi^2_{RSD}+ \chi^2_{CC}+\chi^2_{HST}+\chi^2_{CMB}+\chi^2_{WL}$.
We modify the code CAMB \cite{Lewis:2002ah} which is freely available and here we implement a numerical algorithm. This numerical algorithm is called to solve the background equations and after that corresponding to each data set we calculate the $\chi^2_{tot}$ values. Finally, we call another code known as \texttt{cosmomc}, a markov chain monte carlo package together with a convergence diagnostic by Gelman-Rubin \cite{Gelman-Rubin} that is used  to extract the cosmological parameters. 
The parameters space for our present model is,
$\mathcal{P}_1\equiv\{\Omega_{c}h^2, \Omega_bh^2, 100 \theta_{MC}, \tau, e_{\pi}, w_x, \xi, n_s, \log[10^{10}A_s]\}$ (nine-dimensional space). Here, $\Omega_c h^2$ is the cold dark matter density, $\Omega_bh^2$ is the baryon density, $100 \theta_{MC}$ is the 
ratio of sound horizon to the angular diameter distance, $\tau$ is the optical depth, $n_s$ is the scalar spectra index, $A_s$ is the amplitude of the initial power spectrum and the remaining $e_{\pi}$, $w_x$, $\xi$ are the model parameters described earlier. 
Certainly, the inclusion of both the interaction rate (in terms of the coupling strength $\xi$) and the parameter $e_{\pi}$ quantifying the anisotropic stress,
extends the parameters space compared to the minimum number of parameters  in $\Lambda$CDM, see \cite{Barrow:2014opa} for a detailed discussion.  Finally, we note that for stable perturbations, one needs to impose $c_{sx}^2 \geq 0$. Here, throughout the analysis we have assumed $c_{sx}^2 =1$. In this connection, we mention that since $w_c =0$ (for CDM), thus, $c_{sc}^2 =0$. 

The priors of specific model parameters have been displayed in Table \ref{tab:priors}.

\begin{table}
\caption{The table displays the flat priors on the cosmological parameters used in this work.}
\begin{center}
\begin{tabular}{c|c}
Parameter                    & Prior\\
\hline
\hline
$\Omega_{c}h^2$              & $[0.01, 0.99]$ \\
$\Omega_{b} h^2$             & $[0.005,0.1]$\\
$100\theta_{MC}$             & $[0.5,10]$\\
$\tau$                       & $[0.01,0.8]$\\
$n_s$                        & $[0.5, 1.5]$\\
$\log[10^{10}A_{s}]$         & $[2.4,4]$\\
$w_x$                        & $[-2, 0]$\\
$e_{\pi}$                    & $[-1,1]$\\
$\xi$                        & $[0,2]$\\
\hline
\end{tabular}
\end{center}
\label{tab:priors}
\end{table}

\begingroup
\squeezetable
\begin{center}
\begin{table*}
\caption{68\% and 95\% confidendence-level constraints on the model parameters of the interacting scenario with anisotropic stress using different combined analyses of the observtaional data. Here, $\Omega_{m0} = \Omega_{c0}+\Omega_{b0}$.}
\begin{tabular}{ccccccccccccccccc}
\hline\hline
Parameters & CMB & CMB+BAO+RSD~ & CMB+BAO+RSD+HST &~ CMB+BAO+WL+HST &~ $\begin{array}[c]{c}\text{CMB+BAO+RSD}\\+ \mbox{WL+HST+JLA+CC} \end{array}$\\ \hline

$\Omega_c h^2$ &
$    0.1202_{-    0.0029-    0.0046}^{+    0.0018+    0.0062}$ &
$    0.1164_{-    0.0018-    0.0054}^{+    0.0030+    0.0048}$ &
$    0.1195_{-    0.0023-    0.0040}^{+    0.0020+    0.0043}$ &
$    0.1232_{-    0.0053-    0.0074}^{+    0.0025+    0.0094}$ &
$    0.1201_{-    0.0030-    0.0045}^{+    0.0018+    0.0053}$
\\

$\Omega_b h^2$ &
$    0.02227_{-  0.00016 -    0.00033}^{+   0.00017 +    0.00032}$     &
$    0.02233_{-    0.00017- 0.00029}^{+  0.00015 +  0.00031}$ &
$    0.02229_{-    0.00016-    0.00029}^{+    0.00017+    0.00029}$ &
$    0.02227_{-    0.00014-    0.00027}^{+    0.00014+    0.00029}$ &
$    0.02228_{-    0.00014-    0.00030}^{+    0.00016+    0.00027}$
\\

$100\theta_{MC}$ &
$    1.04048_{-    0.00037-    0.00077}^{+    0.00037+    0.00071}$ &
$    1.04074_{-    0.00037-    0.00068}^{+    0.00038+    0.00071}$ &
$    1.04052_{- 0.00033- 0.00065}^{+    0.00034+    0.00063}$ &
$    1.04031_{-    0.00036-    0.00087}^{+    0.00040+    0.00086}$ &
$    1.04052_{-    0.00034-    0.00071}^{+    0.00037+    0.00072}$
\\

$\tau$ &
$    0.0656_{-    0.0213-    0.0391}^{+    0.0199+    0.0399}$ &
$    0.0707_{-    0.0207-    0.0405}^{+    0.0220+    0.0380}$ &
$    0.0648_{-    0.0181-    0.0370}^{+    0.0186+    0.0347}$ &
$    0.0731_{-    0.0183-    0.0335}^{+    0.0174+    0.0347}$ &
$    0.0653_{-    0.0170-    0.0354}^{+    0.0168+    0.0326}$
\\

$n_s$ &
$    0.9743_{-    0.0045-    0.0090}^{+    0.0046+    0.0086}$ &
$    0.9769_{-    0.0045-    0.0085}^{+    0.0049+    0.0082}$ &
$    0.9751_{-    0.0041-    0.0077}^{+    0.0041+    0.0077}$ &
$    0.9749_{-    0.0038-    0.0074}^{+    0.0039+    0.0076}$ &
$    0.9755_{-    0.0039-    0.0075}^{+    0.0039+    0.0073}$
\\

${\rm{ln}}(10^{10} A_s)$ &
$    3.0741_{-    0.0409-    0.0788}^{+    0.0397+    0.0786}$ &
$    3.0806_{-    0.0409-    0.0809}^{+    0.0439+    0.0770}$ &
$    3.0700_{-    0.0361-    0.0726}^{+    0.0396+    0.0678}$ &
$    3.0882_{-    0.0332-    0.0659}^{+    0.0340+    0.0676}$ &
$    3.0706_{-    0.0329-    0.0674}^{+    0.0356+    0.0639}$
\\

$e_{\pi}$ &
$    0.0852_{-    0.0469-    0.0905}^{+    0.0569+    0.0812}$ &
$    0.0586_{-    0.0692-    0.0853}^{+    0.0733+    0.0981}$ &
$    0.0361_{-    0.0477-    0.0713}^{+    0.0314+    0.0802}$ &
$   -0.0014_{-    0.0248-    0.0494}^{+    0.0237+    0.0518}$ &
$   -0.0064_{-    0.0277-    0.0423}^{+    0.0194+    0.0515}$
\\

$w_x$ &
$   -1.0445_{-    0.1373-    0.4093}^{+    0.1967+    0.3800}$ &
$   -0.9494_{-    0.0415-    0.0838}^{+    0.0392+    0.0832}$ &
$   -1.0349_{-    0.0437-    0.0753}^{+    0.0351+    0.0842}$ &
$   -1.1077_{-    0.0455-    0.0956}^{+    0.0488+    0.0907}$ &
$   -1.0452_{-    0.0276-    0.0755}^{+    0.0408+    0.0626}$
\\

$\xi$ &
$    0.0895_{-    0.0895-    0.0895}^{+    0.0317+    0.2348}$ &
$    0.0931_{-    0.0882-    0.0931}^{+    0.0265+    0.1311}$ &
$    0.0829_{-    0.0829-    0.0829}^{+    0.0261+    0.1017}$ &
$    0.1343_{-    0.1343-    0.1343}^{+    0.0233+    0.2251}$ &
$    0.1119_{-    0.1119-    0.1119}^{+    0.0206+    0.1831}$
\\

$\Omega_{m0}$ &
$    0.3088_{-    0.0407-    0.1002}^{+    0.0375+    0.1015}$ &
$    0.3160_{-    0.0088-    0.0169}^{+    0.0094+    0.0169}$ &
$    0.3036_{-    0.0077-    0.0148}^{+    0.0076+    0.0153}$ &
$    0.2962_{-    0.0140-    0.0253}^{+    0.0112+    0.0265}$ &
$    0.3021_{-    0.0074-    0.0152}^{+    0.0076+    0.0144}$
\\

$\sigma_8$ &
$    0.8489_{-    0.0374-    0.0932}^{+    0.0346+    0.0925}$ &
$    0.8266_{-    0.0161-    0.0281}^{+    0.0132+    0.0302}$ &
$    0.8242_{-    0.0143-    0.0293}^{+    0.0145+    0.0275}$ &
$    0.8204_{-    0.0231-    0.0608}^{+    0.0318+    0.0545}$ &
$    0.8116_{-    0.0144-    0.0335}^{+    0.0188+    0.0289}$
\\

$H_0$ &
$   68.6361_{-    5.7559-   10.9049}^{+    4.0664+   12.5331}$ &
$   66.4256_{-    1.0226-    1.7988}^{+    0.9544+    1.8935}$ &
$   68.5197_{-    0.9289-    1.9400}^{+    1.0257+    1.7763}$ &
$   70.2674_{-    1.2938-    2.2335}^{+    1.0577+    2.3953}$ &
$   68.8154_{-    0.9007-    1.6168}^{+    0.7117+    1.8013}$
\\

\hline\hline
\end{tabular}
\label{tab:results}
\end{table*}
\end{center}
\endgroup

\begin{figure*}
\includegraphics[width=0.9\textwidth]{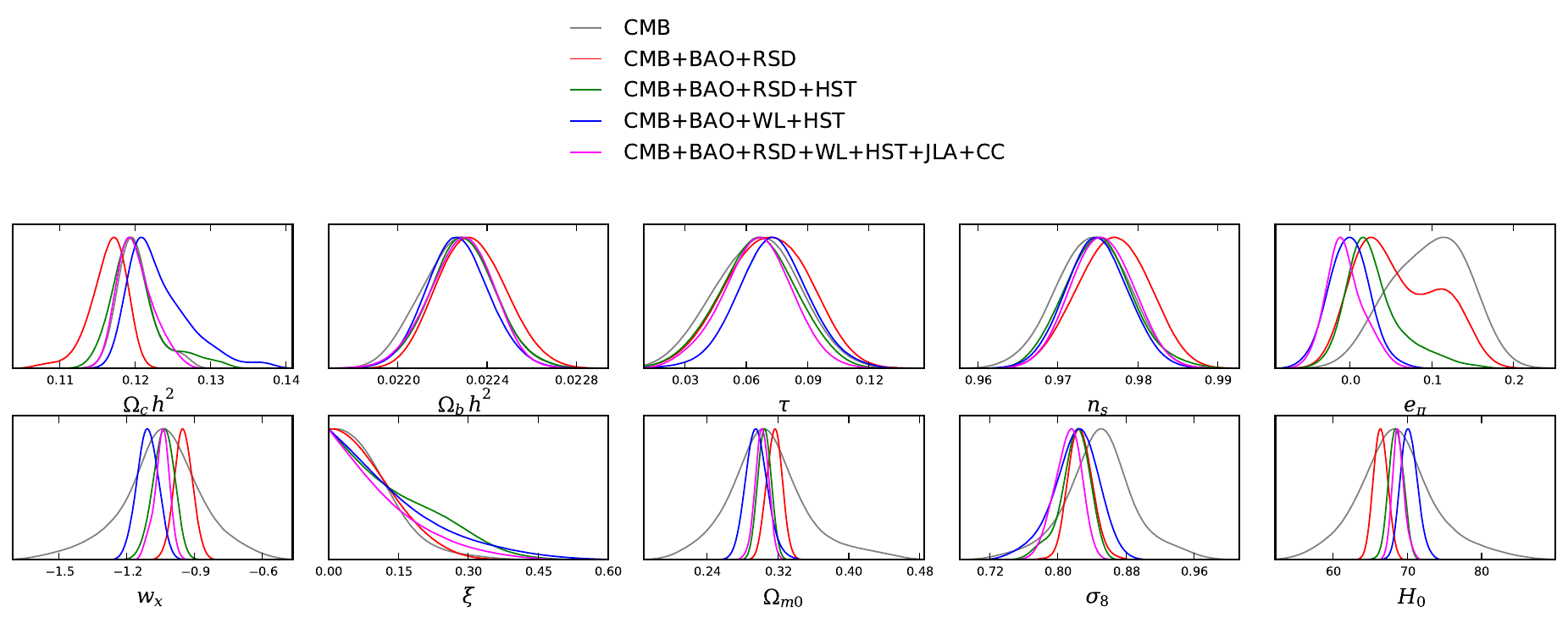}
\caption{The plots show the one-dimensional posterior distributions for various cosmological parameters using different combined analysis of the observational data as displayed in Table \ref{tab:results}.}
\label{fig:posterior}
\end{figure*}

\begin{figure*}
\includegraphics[width=0.32\textwidth]{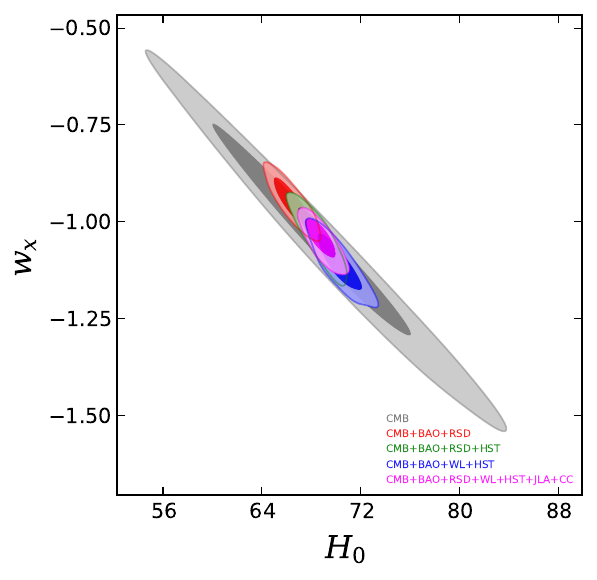}
\includegraphics[width=0.325\textwidth]{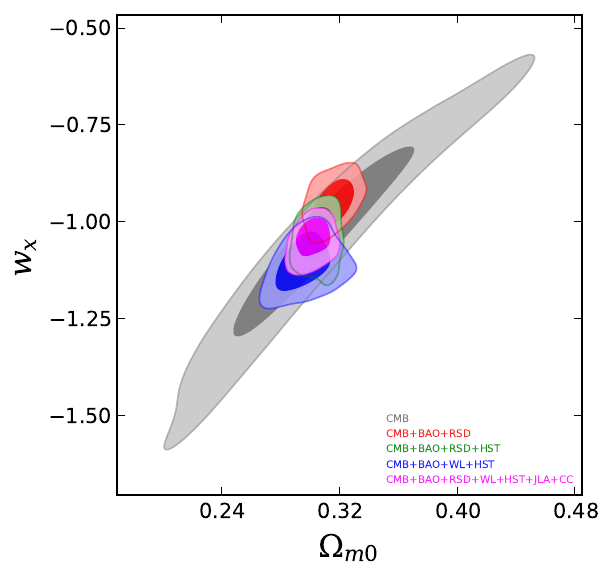}
\includegraphics[width=0.325\textwidth]{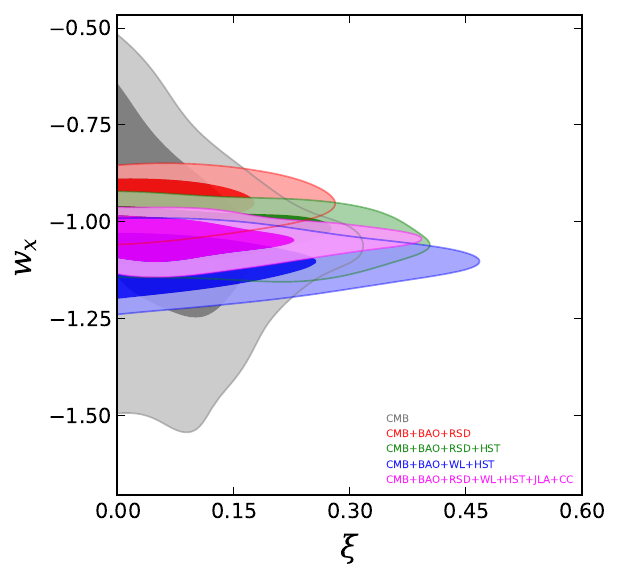}
\caption{68\% and 95\% confidence-level contour plots in the two-dimensional $(H_0, w_x)$, $(\Omega_{m0}, w_x)$ and $(\xi, w_x)$ planes for different combined analyses have been shown. \texttt{Left Panel:} This shows that higher values of $H_0$ allow more phantom nature in the dark energy equation of state $w_x$, while the quintessence nature is favoured in $w_x$ for lower values of $H_0$. \texttt{Middle Panel:} Higher values of $\Omega_{m0}$ favor the quintessence character in the dark energy equation of state while the phantom character of $w_x$ is increased with the lower values of $\Omega_{m0}$. \texttt{Right panel:} The parameters $w_x$ and $\xi$ are almost uncorrelated with each other. }
\label{fig:1}
\end{figure*}

\section{Results and analysis}
\label{sec-results}

The interacting scenario in presence of the matter-sourced anisotropic stress is the main focus of the work. However, we have also constrained 
the interacting scenario where no matter sourced anisotropic stress is present. The motivation of the second analysis is to see how the presence of matter sourced anisotropic stress affects the cosmological dynamics. In order to  
constrain both the interacting scenarios we have used the following observational data: 
\begin{enumerate}
\item CMB (Planck TTTEEE+lowTEB), 

\item CMB+BAO+RSD,  

\item CMB+BAO+RSD+HST,  

\item CMB+BAO+WL+HST, 

\item CMB+BAO+RSD+WL+HST+JLA+CC. 
\end{enumerate}

For the interacting scenario with matter sourced anisotropic stress, we have presented the observational summary in Table \ref{tab:results} where the constraints on the model parameters are shown at 68\% and 95\% confidence levels. In Fig. \ref{fig:posterior}, we display the one-dimensional posterior distributions for some selected model parameters for the above observational data. Let us now analyze the observational constraints on the model parameters.

From Table \ref{tab:results} we see that the observational data  allow a non-zero interaction between dark matter and dark energy. However, observing the $1\sigma$ error-bars of the coupling parameter, $\xi$, one can readily conclude that $\xi =0$ is allowed by almost all observational data. That means within $1\sigma$ confidence-level, a non-interacting $w$CDM model is still allowed. From the the dark energy equation of state we find that only CMB and the combined analysis CMB+BAO+RSD hint for its quintessence nature. One can see that the CMB data alone constrain the dark energy equation of state, $w_x = -1.0445_{-    0.1373}^{+    0.1967}$ at 68\% CL ($-1.0445_{- 0.4093}^{+ 0.3800}$ at 95\% CL) while from the combination CMB+BAO+RSD, we find $w_x= -0.9494_{- 0.0415}^{+ 0.0392}$ at 68\% CL ($-0.9494_{- 0.0838}^{+ 0.0832}$ at 95\% CL). One may notice that the addition of BAO and RSD to CMB decreases the error bars in the dark energy equation of state, that means the parameter space for $w_x$ gets reduced. 
Interestingly enough, when the $H_0$ prior from the HST is included to the other data sets, the dark energy equation of state moves toward the cosmological constant boundary. The combined analysis CMB+BAO+RSD+HST shows that $w_x= -1.0349_{-    0.0437}^{+    0.0351}$ at 68\% CL ($-1.0349_{- 0.0753}^{+0.0842}$ at 95\% CL). The last two analyses, namely,
CMB+BAO+WL+HST and CMB+BAO+RSD+WL+HST+CC+JLA infer the same about the dark energy equation of state, see the last two columns of Table \ref{tab:results}.
Thus, from the results, one can identify that the dark energy sector resembles with the cosmological constant. Hence, one may conclude that, although a non-zero deviation from the $\Lambda$CDM cosmology is favored by the observational data but effectively, such deviation is very minimal and hence the model is close to the $\Lambda$CDM model. In Fig. \ref{fig:1} we display the 68\% and 95\% confidence-level contour plots for the combinations ($w_x$, $H_0$), ($w_x$, $\Omega_{m0}$) and ($w_x$, $\xi$) using different combined analyses performed in this work. From the left panel of Fig. \ref{fig:1}, we find that for lower values of $H_0$, the dark energy equation of state, $w_x$ has a shifting nature towards the quintessence regime while from the middle panel of this figure, we observe that, as $w_x$ increases, that means when it shifts towards the quintessence regime, the density parameter for cold dark matter increases. From the right panel of Fig. \ref{fig:1}, we show the dependence of coupling parameter $\xi$ with the dark energy equation of state, $w_x$, from which making any decisive conclusion between the dependence of $\xi$ with $w_x$ looks very hard, in fact, the parameters $w_x$ and $\xi$ look uncorrelated  with each other.

\begin{figure*}
\includegraphics[width=0.32\textwidth]{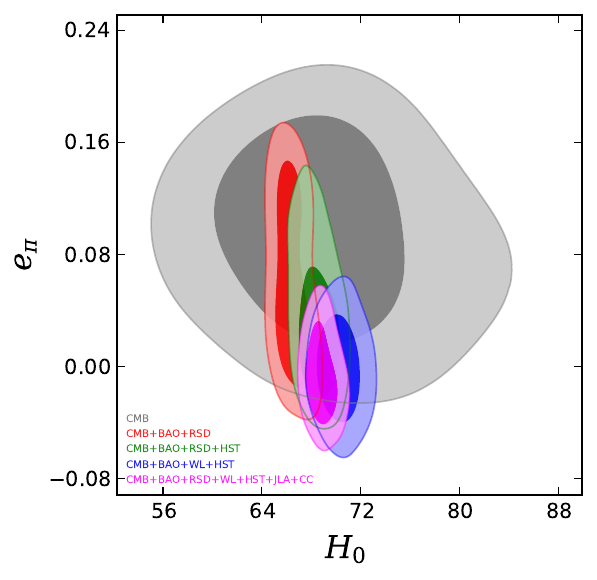}
\includegraphics[width=0.326\textwidth]{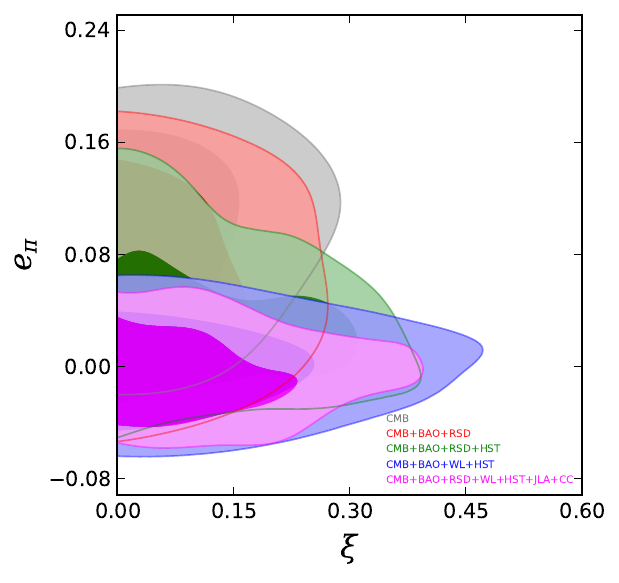}
\includegraphics[width=0.32\textwidth]{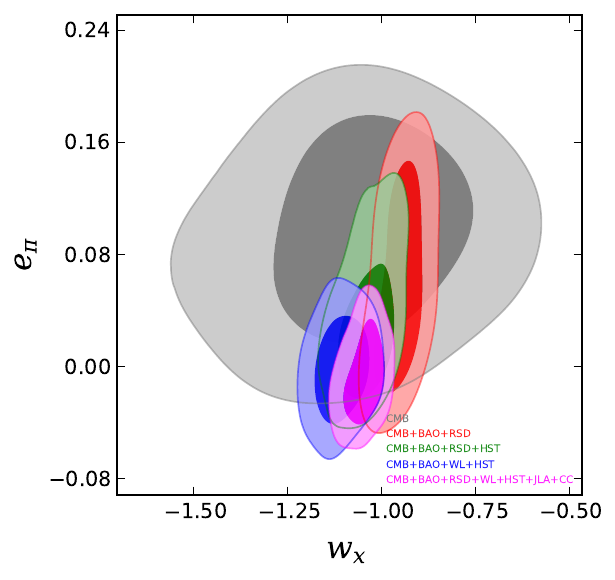}
\caption{68\% and 95\% confidence-level contour plots in $(e_{\pi}, H_0)$, $(e_{\pi}, \xi)$, and $(e_{\pi}, w_x)$ planes  have been shown for several observational combinations. \texttt{Left Panel:} This shows the $(H_0, e_{\pi})$ plane. One can see that the combination CMB+ext (where `ext' is the other data sets, for instance BAO, RSD,.. etc) decreases the error bars on the parameters. Although, one cannot find a clear relation between $e_{\pi}$ and the Hubble parameter values, but the plots for different combinations (except CMB) slightly show that $e_{\pi}$ has a very weak tendency to increase its values for lower values of $H_0$. We repeat that such tendency is extremely weak according to the current data we employ. \texttt{Middle Panel:} This actually infers a low interaction scenario with a small anisotropic stress. However, one can clearly notice that the parameters ($e_{\pi}$, $\xi$) are almost uncorrelated with each other. \texttt{Right Panel:} One can see that the phantom dark energy allows lower values of $e_{\pi}$ while for quintessence dark energy one may expect slightly higher values of $e_{\pi}$, although, it is clear that the observational data do not allow a large $e_{\pi}$. }
\label{fig:2}
\end{figure*}

\begin{figure*}
\includegraphics[width=0.32\textwidth]{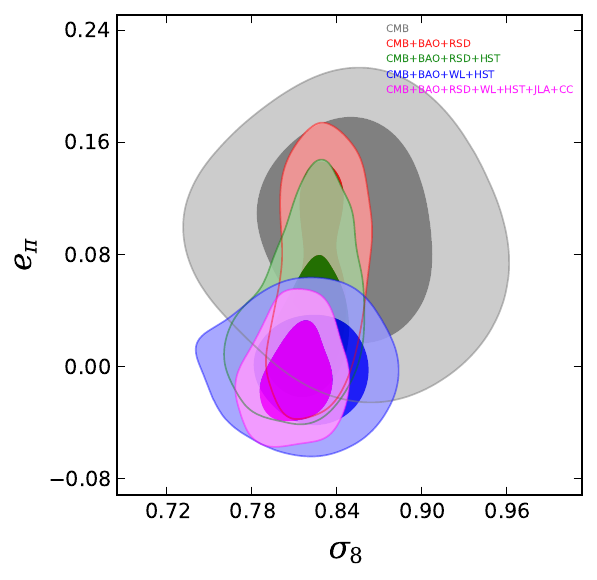}
\includegraphics[width=0.32\textwidth]{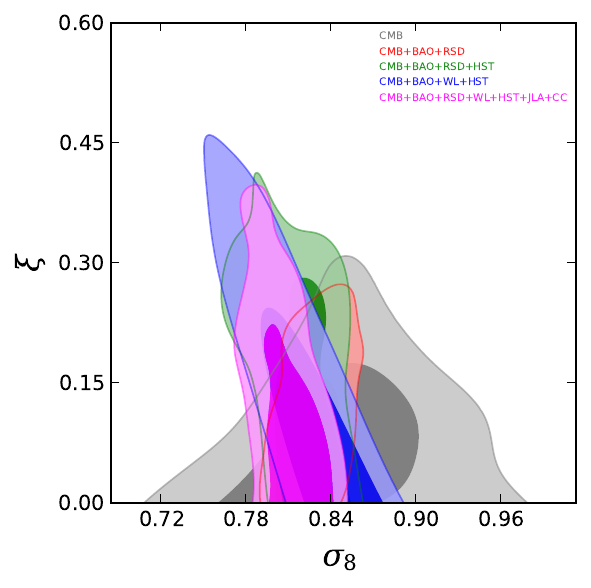}
\includegraphics[width=0.318\textwidth]{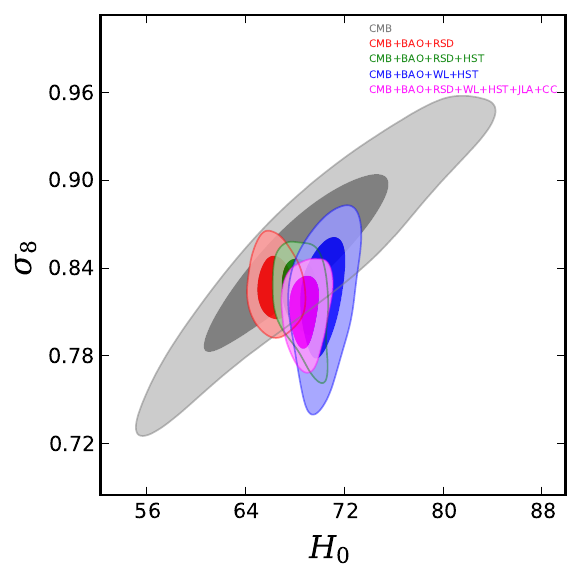}
\caption{68\% and 95\% confidence-level contour plots in ($\sigma_8$, $e_{\pi}$), ($\sigma_8$, $\xi$) and ($H_0$, $\sigma_8$) planes for several observational combinations. \texttt{Left Panel:} From the plot, we do not observe any significant effect on 
$\sigma_8$ for anisotropic stress. In fact, one may see that a small value of $\pi$ is allowed in agreement with the estimated value of $\sigma_8$ from Planck Ade et al. (2016). \texttt{Middle Panel:} One may notice that $\sigma_8$ has a slight dependence on $\xi$, although such dependence is weak but this is not null. One can see that $\sigma_8$ has a tendency to take lower values for increasing strength of the interaction. \texttt{Right Panel:} A weakly dependence  between $H_0$ and $\sigma_8$ is reflected from this plot. }
\label{fig:3}
\end{figure*}

The inclusion of $H_0$ prior from HST also affects other cosmological parameters. For instance, from the constraints on the anisotropic stress displayed in Table \ref{tab:results} one can find the considerable changes in its constraints. The magnitude of the anisotropic stress significantly changes. The only CMB data constrain $e_{\pi} = 0.0852_{- 0.0469}^{+ 0.0569}$ at 68\% CL ($ 0.0852_{-0.0905}^{+ 0.0812}$ at 95\% CL) while from  CMB+BAO+WL+HST, $ e_{\pi} = -0.0014_{- 0.0248}^{+ 0.0237}$ at 68\% CL ($ -0.0014_{- 0.0494}^{+ 0.0518}$ at 95\% CL) and from  the full combination CMB+BAO+RSD+WL+HST+JLA+CC, it is $e_{\pi}= -0.0064_{- 0.0277}^{+ 0.0194}$ at 68\% CL ($ -0.0064_{-0.0423}^{+0.0515}$ at 95\% CL). One may observe that there is no such significant changes in the error bars in the anisotropic stress.  In Fig. \ref{fig:2}, we present the 68\% and 95\% confidence-level contour plots where we show the effects of the anisotropic stress on some selected cosmological parameters, namely, $H_0$, $\xi$ and $w_x$. Additionally, in Fig. \ref{fig:3} we show the $\sigma_8$ dependence on other cosmological parameters, namely, $e_{\pi}$, $\xi$ and $H_0$.

\begin{figure*}
\includegraphics[width=0.45\textwidth]{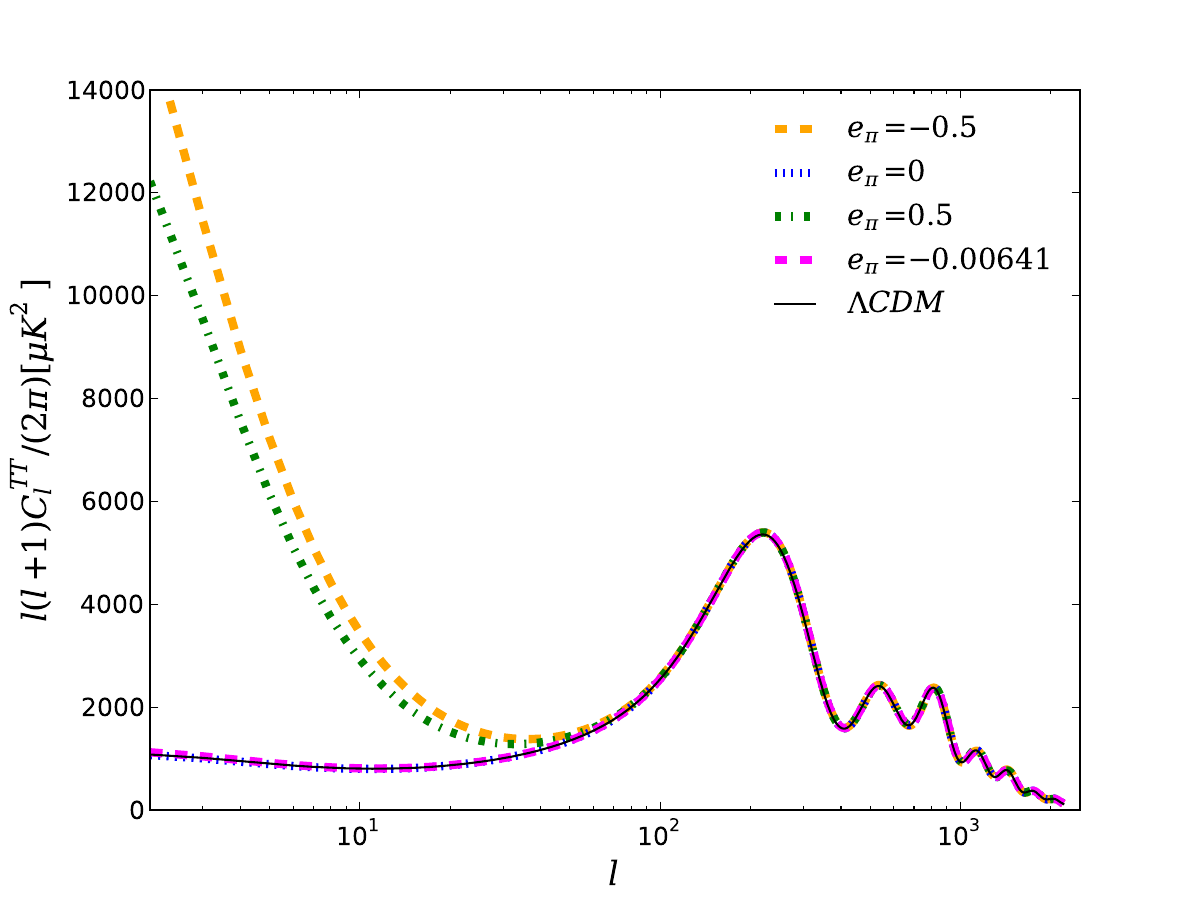}
\includegraphics[width=0.45\textwidth]{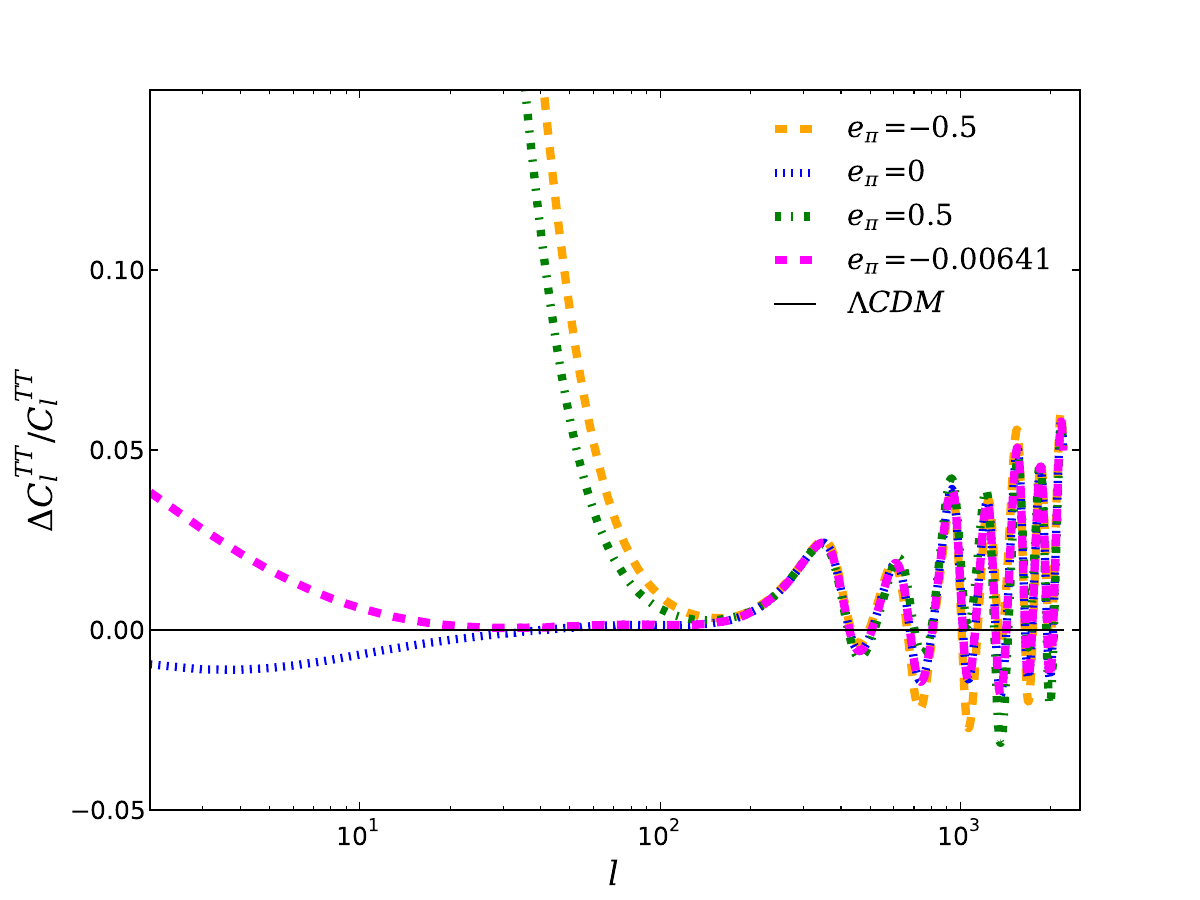}
\caption{The figure shows the CMB TT power spectra (Left Panel) and the ratio (also known as the relative deviation) of the CMB TT power spectra (Right Panel) for  the present interacting dark-energy scenario $Q = 3 H \xi (1+w_x) \rho_x$ with and without the presence of anisotropic stress that we consider in this work (see section \ref{sec-bg+per} for details). 
Here, $\Delta C_l^{TT} =C_{l}^{TT}\bigl|_{model} \, -\, C_{l}^{TT}\bigl|_{LCDM}$ and $C_l^{TT} = C_{l}^{TT}\bigl|_{LCDM}$. From the Left Panel, one may notice that at low angular scales, with the increase of $|e_{\pi}|$, the deviation from the non-interacting $\Lambda$CDM becomes high, but however, at high angular scales, no such deviation in the CMB TT spectra for $|e_{\pi}|$ is observed. The similar behaviour is reflected from the Right Panel, although a non-zero deviation from the $\Lambda$CDM even at high angular scales is observed here. }
\label{fig:cmbTT1}
\end{figure*}

\begin{figure*}
\includegraphics[width=0.45\textwidth]{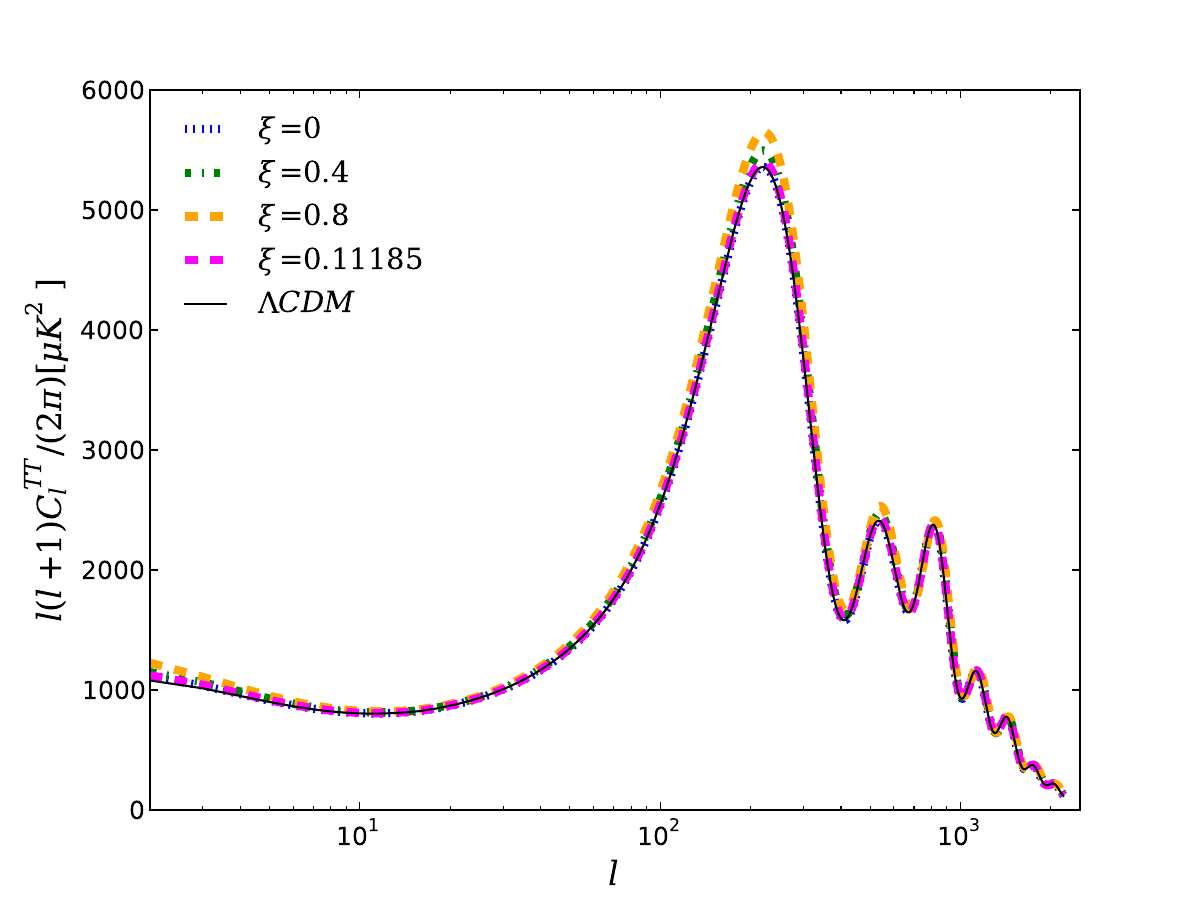}
\includegraphics[width=0.45\textwidth]{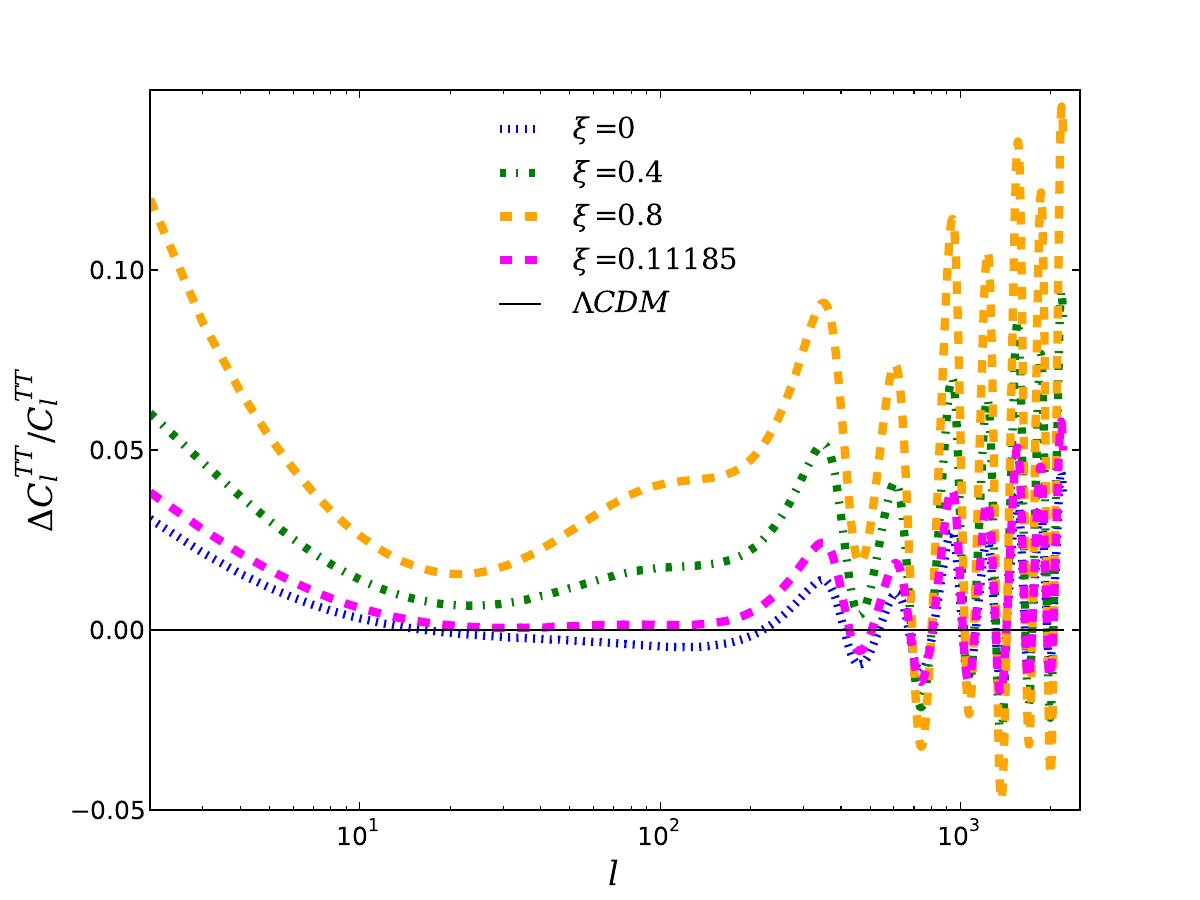}
\caption{The figure shows the CMB TT power spectra (Left Panel) and the ratio of the CMB TT power spectra (Right Panel) for the interacting dark-energy scenario in presence of the anisotropic stress  considered in this work (described in section \ref{sec-bg+per}). From the Left Panel we notice that even in presence of an anisotropic stress sourced by the matter field, the deviation in the CMB TT spectra mainly appears due to large values of the coupling parameter. The Right Panel confirms the observation of the Left Panel and additionally a deviation from the $\Lambda$CDM. }
\label{fig:cmbTT2}
\end{figure*}

We now focus on the dynamics of the univese on the large scales for the current cosmological scenario.  In Fig. \ref{fig:cmbTT1}, we have plotted the CMB TT power spectra (see the left panel of Fig. \ref{fig:cmbTT1}) and the ratio of the CMB TT power spectra (see the right panel of Fig. \ref{fig:cmbTT1}) for different values of the anisotropic stess $e_{\pi}$ and compared the analyses with the base $\Lambda$CDM model. It is quite clear from this figure that at low angular scales, for large anisotropic stress, the model deviates vastly from the $\Lambda$CDM model while as the angular scale increases, the deviation reduces from the $\Lambda$CDM and at high angular scales, the anisotropic stress does not produce any effective changes in the power spectra. However, the right panel of Fig. \ref{fig:cmbTT1} says something more which is not visible from the left panel of Fig. \ref{fig:cmbTT1}. From the ratio of CMB TT spectra displayed in the right panel of Fig. \ref{fig:cmbTT1}, one can see that the model still shows a slight deviation from $\Lambda$CDM in small angular scales and even if a nonzero value of the anisotropic stress is allowed. Thus, the model has a slight difference from $\Lambda$CDM and such a difference is  very small.  However, we have a very interesting observation from  Fig. \ref{fig:cmbTT2} displaying the CMB TT spectra and the ratio of the CMB TT spectra for different strengths of the coupling
parameters. From the left panel of Fig. \ref{fig:cmbTT2}  one can see that the a slight deviation of the stressed interacting scenario from the $\Lambda$CDM model is observed for a large value of the coupling parameter ($\xi = 0.8$, which is a very big value in compared to the observational estimation summarized in Table \ref{tab:results}) while from the right panel of Fig. \ref{fig:cmbTT2}, it is quite clear that the model definitely has a deviation from the base $\Lambda$CDM for any $\xi \neq 0$. However, the deviation is not much significant.

\begingroup
\squeezetable
\begin{center}
\begin{table*}
\caption{68\% and 95\% confidendence-level constraints on the model parameters of the interacting scenario with no anisotropic stress for different combined analyses. Here, $\Omega_{m0} = \Omega_{c0}+\Omega_{b0}$.}
\begin{tabular}{ccccccccccccccccc}
\hline\hline
Parameters & CMB & CMB+BAO+RSD & CMB+BAO+RSD+HST & CMB+BAO+WL+HST & $\begin{array}[c]{c}\text{CMB+BAO+RSD}\\+ \mbox{WL+HST+JLA+CC} \end{array}$\\ \hline

$\Omega_c h^2$ &
$    0.1214_{-    0.0032-    0.0046}^{+    0.0024+    0.0052}$ &
$    0.1180_{-    0.0014-    0.0051}^{+    0.0025+    0.0042}$ &
$    0.1206_{-    0.0043-    0.0064}^{+    0.0019+    0.0085}$ &
$    0.1190_{-    0.0022-    0.0041}^{+    0.0020+    0.0044}$ &
$    0.1183_{-    0.0014-    0.0029}^{+    0.0014+    0.0030}$
\\

$\Omega_b h^2$ &
$    0.02220_{-    0.00016-    0.00031}^{+    0.00016+    0.00029}$ &
$    0.02223_{-    0.00014-    0.00029}^{+    0.00015+    0.00030}$ &
$    0.02229_{-    0.00015-    0.00027}^{+    0.00015+    0.00029}$ &
$    0.02226_{-    0.00014-    0.00030}^{+    0.00016+    0.00027}$ &
$     0.02231_{-    0.00014-    0.00029}^{+    0.00014+    0.00029}$
\\

$100\theta_{MC}$ &
$    1.04038_{-    0.00034-    0.00065}^{+    0.00033+    0.00064}$ &
$    1.04055_{-    0.00037-    0.00063}^{+    0.00032+    0.00066}$ &
$    1.04058_{-    0.00032-    0.00063}^{+    0.00031+    0.00063}$ &
$    1.04040_{-    0.00036-    0.00074}^{+    0.00041+    0.00069}$ &
$     1.04065_{-    0.00033-    0.00063}^{+    0.00034+    0.00060}$
\\

$\tau$ &
$    0.0778_{-    0.0167-    0.0336}^{+    0.0166+    0.0326}$ &
$    0.0674_{-    0.0180-    0.0373}^{+    0.0196+    0.0371}$ &
$    0.0704_{-    0.0197-    0.0371}^{+    0.0186+    0.0368}$ &
$0.0748_{-    0.0176-    0.0328}^{+    0.0168+    0.0359}$ &
$0.0663_{-    0.0162-    0.0319}^{+    0.0161+    0.0315}$
\\

$n_s$ &
$    0.9729_{-    0.0044-    0.0086}^{+    0.0047+    0.0088}$ &
$0.9742_{-    0.0039-    0.0087}^{+    0.0047+    0.0085}$ &
$    0.9754_{-    0.0045-    0.0083}^{+    0.0044+    0.0086}$ &
$0.9751_{-    0.0043-    0.0083}^{+    0.0041+    0.0080}$ &
$0.9760_{-    0.0038-    0.0070}^{+    0.0036+    0.0071}$\\

${\rm{ln}}(10^{10} A_s)$ &
$    3.0993_{-    0.0336-    0.0675}^{+    0.0330+    0.0692}$ &
$3.0764_{-    0.0339-    0.0725}^{+    0.0382+    0.0710}$ &
$    3.0822_{-    0.0385-    0.07034}^{+    0.0369+    0.0698}$ &
$3.0906_{-    0.0341-    0.0636}^{+    0.0330+    0.0689}$ &
$3.0722_{-    0.0288-    0.0616}^{+    0.0311+    0.0605}$
\\

$w_x$ &
$   -1.0737_{-    0.0997-    0.3034}^{+    0.134+    0.2517}$ &
$   -0.9636_{-    0.0595-    0.0823}^{+    0.0333+    0.1058}$ &
$   -1.0282_{-    0.0559-    0.0802}^{+    0.0406+    0.0864}$ &
$-1.1040_{-    0.0462-    0.0893}^{+    0.0498+    0.0909}$ &
$-1.0230_{-    0.0257-    0.0603}^{+    0.0329+    0.0527}$
\\

$\xi$ &
$    0.1372_{-    0.1292-    0.1372}^{+    0.0382+    0.1915}$ &
$    0.0849_{-    0.0849-    0.0849}^{+    0.0222+    0.1598}$ &
$    0.0577_{-    0.0577-    0.0577}^{+    0.0170+    0.0998}$ &
$ 0.0849_{-    0.0849-    0.0849}^{+    0.0209+    0.1024}$ &
$0.0360_{-    0.0360-    0.0360}^{+    0.0091+    0.0507}$
\\

$\Omega_{m0}$ &
$    0.3045_{-    0.0251-    0.0646}^{+    0.0279+    0.0599}$ &
$    0.3205_{-    0.0118-    0.0187}^{+    0.0086+    0.0227}$ &
$    0.3040_{-    0.0083-    0.0163}^{+    0.0083+    0.0167}$ &
$0.2925_{-    0.0075-    0.0163}^{+    0.0088+    0.0147}$ &
$0.3014_{-    0.0077-    0.0141}^{+    0.0070+    0.0139}$
\\

$\sigma_8$ &
$    0.8395_{-    0.0308-    0.0549}^{+    0.0241+    0.0663}$ &
$0.8120_{-    0.0146-    0.0296}^{+    0.0137+    0.0296}$ &
$    0.8212_{-    0.0166-    0.0255}^{+    0.0133+    0.0287}$ &
$0.8295_{-    0.0179-    0.0368}^{+    0.0191+    0.0385}$ &
$0.8156_{-    0.0137-    0.0244}^{+    0.0121+    0.0246}$
\\

$H_0$ &
$   69.0829_{-    3.9006-    6.8847}^{+    2.7931+    9.0963}$ &
$66.3205_{-    0.9677-    2.6570}^{+    1.4094+    2.1702}$ &
$   68.3578_{-    0.9759-    2.0948}^{+    1.2570+    1.8742}$ &
$70.3302_{-    1.1574-    1.9285}^{+    1.0053+    2.1042}$ &
$68.4646_{-    0.7380-    1.3616}^{+    0.8199+    1.3348}$
\\

\hline\hline
\end{tabular}
\label{tab:results-no-stress}
\end{table*}
\end{center}
\endgroup

\begin{figure*}
\includegraphics[width=0.7\textwidth]{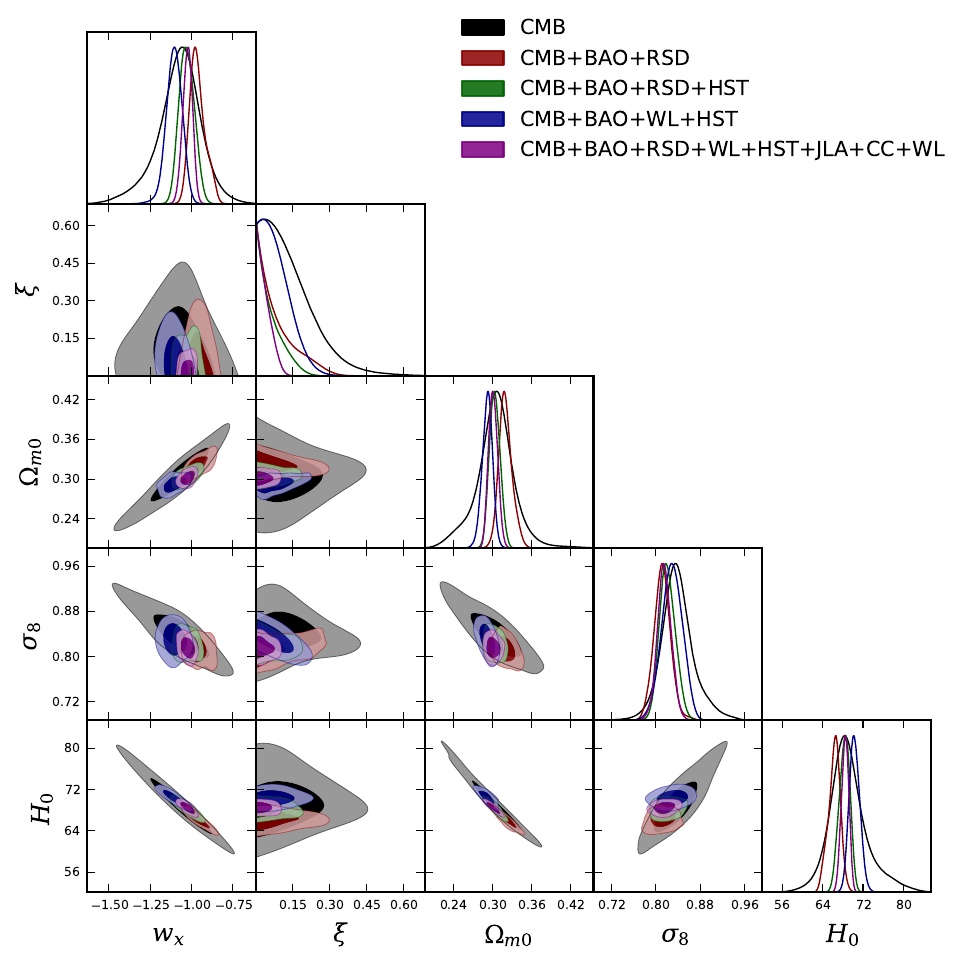}
\caption{68\% and 95\% confidence level contour plots for the interacting sceanrio with no-anisotropic stress have been shown using different combined analysis. The figure also contains the 2-dimensional posterior distributions for the parameters ($w_x$, $\xi$, $\Omega_{m0}$, $\sigma_8$, $H_0$). Here, the parameter $\Omega_{m0}$ is the current value of $\Omega_m = \Omega_c+\Omega_b$. }
\label{contour-no-anisotropy}
\end{figure*}

\begin{figure*}
\includegraphics[width=0.45\textwidth]{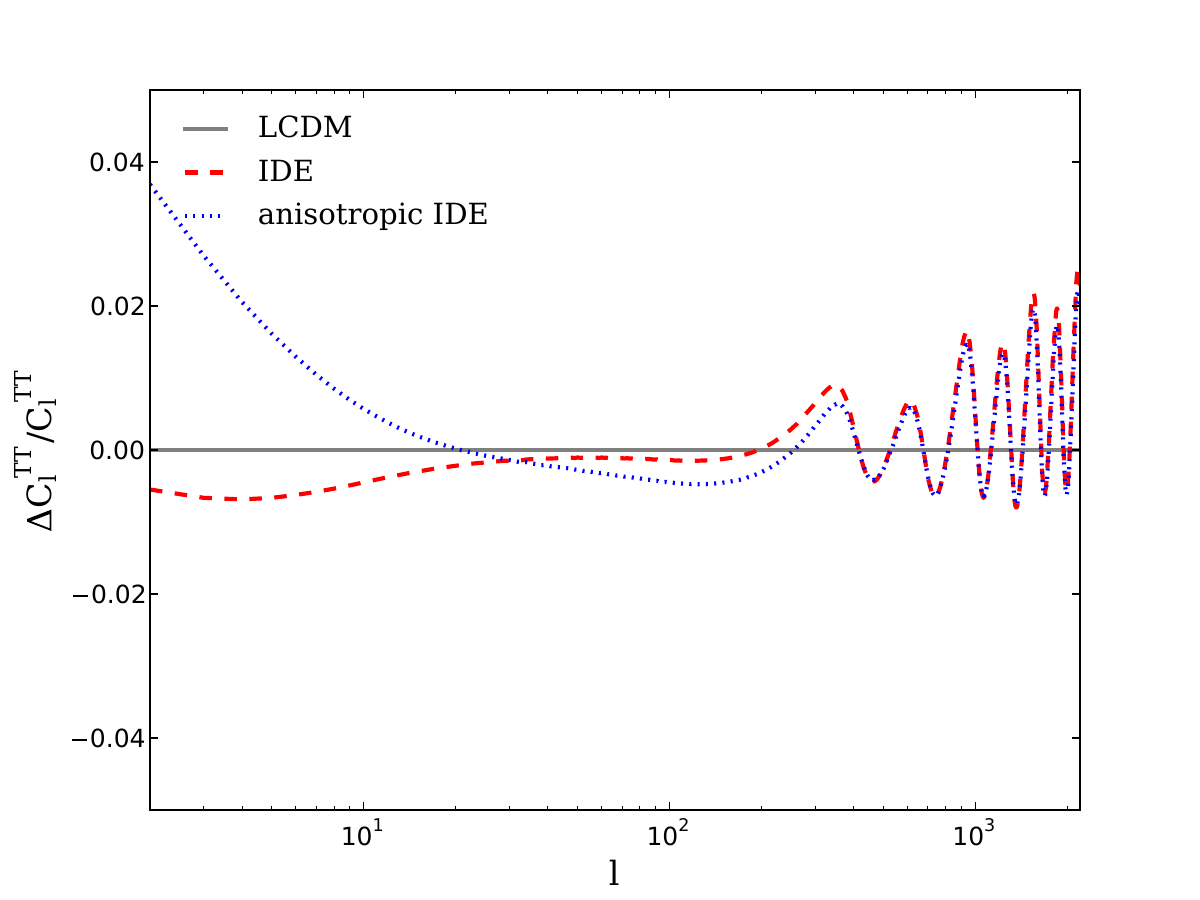}
\includegraphics[width=0.45\textwidth]{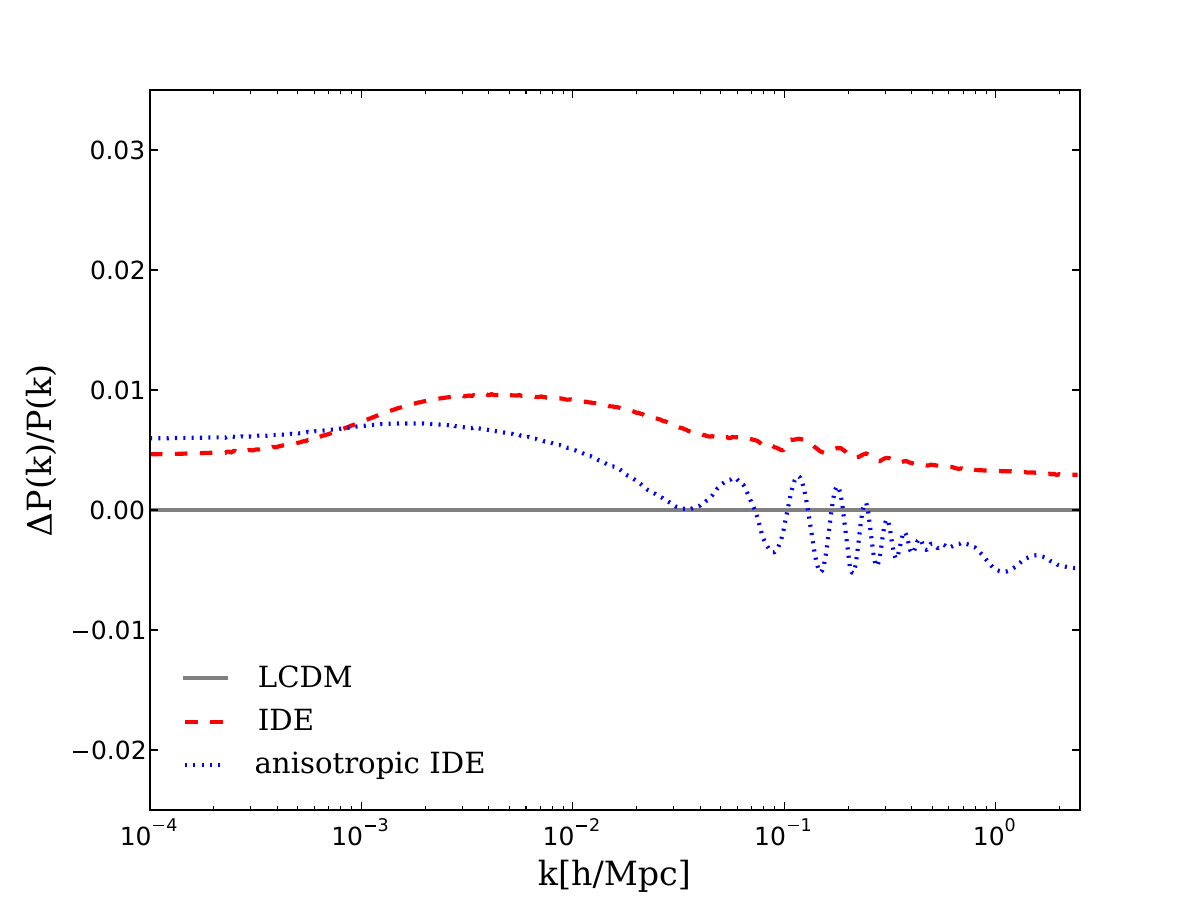}
\caption{The relative deviations in the CMB TT spectra (left panel) and matter power spectra (right panel) have been shown for the interacting scenarios with and without the anisotropic stress using the combined observational data CMB+BAO+RSD+WL+HST+JLA+CC. }
\label{fig:comparison}
\end{figure*}

\subsection{Comparison with no-anisotropic stress}
\label{sec-no-aniostropy}

In the previous section \ref{sec-results} we have studied the effects of the anisotropic stress on the cosmological parameters when the dark fluids are interacting with each other. A question that immediately appears in this context is, how the contribution of anisotropic stress affects the large scale structure of the universe and also in the
estimation of the cosmological parameters? To answer these questions, we perform similar  analyses making $e_{\pi} = 0$ in the evolution equations with the same
priors presented in Table \ref{tab:priors}.
It is quite evident that the analysis with no-anisotropic stress will
effectively present a quantitative and qualitative differences on the cosmological parameters. The  observational constraints for this particular scenario have been shown in Table \ref{tab:results-no-stress}. The 68\% and 95\% confidence level contour plots for some selected parameters
have been presented in Fig. \ref{contour-no-anisotropy} where we also show their one-dimensional posterior distributions. From both the analyses, one can clearly notice that except for CMB data only, the exclusion of the anisotropic stress lowers the strength of the coupling parameter. This is an interesting observation in this work which clearly demonstrates that the addition of $e_{\pi}$ is the measure of increment in the coupling strength, $\xi$.
However, from the analyses presented in Table \ref{tab:results} and Table \ref{tab:results-no-stress}, a clear conclusion that one might draw is, in both the scenarios (with and without anisotropic stress), the coupling strength recovers its zero value within the 68\% CL, that means the model $w_x$CDM$+\xi+e_{\pi}$ may recover the
non-interacting $w_x$CDM cosmology within this 68\% CL and this model has a close resemble with that of the $\Lambda$CDM cosmology. This might be considered to be a common behavior of the models. Furthermore, it should be mentioned that only the combination CMB+BAO+RSD with $e_{\pi} \neq 0$, does not recover the $\xi =0$ limit in anyway and thus, this particular combination always indicates for a non-zero interaction in the dark sector.

Probably the most interesting observation comes from the relative deviation of the CMB TT and matter power spectra shown in Fig. \ref{fig:comparison}. A quick look says that both the scenarios are close to $\Lambda$CDM but there is something more that we would like to describe here. Let us focus on the left panel of Fig. \ref{fig:comparison} where the relative deviation of the CMB TT spectra has been shown. One can notice that for $l\lesssim 10$, the scenario $w_x$CDM$+\xi+e_{\pi}$ significantly differs from $w_x$CDM$+\xi$, and with the increase of $l$ up to a certain value, the difference between the scenarios decreases. We find a  particular value of $l$ residing in the region $10 <l < 10^2$ where the difference between the models becomes zero  (the point at which both the plots intersect with each other), but, after that, up to some certain $l$, the difference between the scenarios again increases where the model $w_x$CDM$+\xi+e_{\pi}$ stays far from $\Lambda$CDM in compared to $w_x$CDM$+\xi$. Further, we again notice that, the model $w_x$CDM$+\xi+e_{\pi}$ approaches toward $\Lambda$CDM and becomes closer in compared to the $w_x$CDM$+\xi$. And for large $l$, both the models seem to be indistinguishable from one another.
We note that, for all $l$, the quantity $\Delta C_l^{TT}/C_l^{TT}$ that reports the difference  of the model from the $\Lambda$CDM is very small.

Now we concentrate on the relative deviation in the matter power spectra  (right panel of Fig. \ref{fig:comparison}), $\Delta P (k)/P(k)$. We remark that for all $k$, the qunatity $\Delta P (k)/P(k)$ is very small informing the closeness of  the interacting scenarios toward the $\Lambda$CDM model, but however, we notice some additional features. We find that for very small $k$,
almost for $k \lesssim 10^{-3}$, the quantity $\Delta P (k)/P(k)$ for the
model $w_x$CDM$+\xi+e_{\pi}$ is slightly greater in compared to the model $w_x$CDM$+\xi$, however, for $k \gtrsim 10^{-3}$, the reverse scenario is observed,  that means, the difference between the interacting scenarios, $w_x$CDM$+\xi+e_{\pi}$ and $w_x$CDM$+\xi$ becomes pronounced. Overall, we notice that both the interacting pictures are close to $\Lambda$CDM, but indeed, they do not overlap with $\Lambda$CDM completely. 

Thus, overall, one may coclude that indeed the scenarios $w_x$CDM$+\xi+e_{\pi}$ and $w_x$CDM$+\xi$ maintain differences amongst each other but for large $l$ (for CMB TT spectra) and large $k$ (matter power spectra), both the scenarios effectively approach toward the  $\Lambda$-cosmology.

\begingroup
\squeezetable
\begin{center}
\begin{table*}
\caption{68\% and 95\% CL constraints on $H_0$ for the models with $e_{\pi} \neq 0$ and $e_{\pi} =0$ for different combined analyses of the observtaional data.}
\begin{tabular}{ccccccccccccccccc}
\hline\hline
Parameters & CMB & CMB+BAO+RSD~ & CMB+BAO+RSD+HST &~ CMB+BAO+WL+HST &~ $\begin{array}[c]{c}\text{CMB+BAO+RSD}\\+ \mbox{WL+HST+JLA+CC} \end{array}$\\ \hline

$H_0$ ($e_{\pi}\neq 0$) &
$   68.64_{-    5.76-   10.90-   12.82}^{+    4.07+   12.53+   17.61}$ &
$66.43_{-    1.02-    1.80-    2.38}^{+    0.95+    1.89+    2.44}$ &
$   68.52_{-    0.93-    1.94-    2.51}^{+    1.03+    1.78+    2.73}$ &
$70.27_{-    1.29-    2.23-    2.67}^{+    1.06+    2.40+    3.25}$ &
$68.82_{-    0.90-    1.62-    1.95}^{+    0.71+    1.80+    2.46}$
\\
$H_0$ ($e_{\pi} =0$) &
$   69.08_{-    3.90-    6.88-   10.20}^{+    2.79+    9.10+   12.90}$ &
$66.32_{-    0.97-    2.66-    2.87}^{+    1.41+    2.17+    2.83}$ &
$   68.36_{-    0.98-    2.09-    2.51}^{+    1.26+    1.87+    2.21}$ &
$70.33_{-    1.16-    1.93-    2.40}^{+    1.01+    2.10+    2.83}$ &
$68.46_{-    0.74-    1.36-    1.87}^{+    0.82+    1.33+    1.66}$
\\

\hline\hline
\end{tabular}
\label{tab:tension}
\end{table*}
\end{center}
\endgroup

\begingroup
\squeezetable
\begin{center}
\begin{table*}
\caption{For different regions of the dark energy state parameter, $w_x$, we constrain the $w_x$CDM$+\xi + e_{\pi}$ scenario using the combined observational data %
CMB+BAO+RSD+WL+HST+JLA+CC. The table shows the mean values of the cosmological parameters with their errors at 68\% and 95\% confidence-levels. }
\begin{tabular}{cccccccccccccc}
\hline\hline
Parameters & $w_x \in [-2, -1.2]$ & $w_x \in [-2, -1]$ & $w_x \in [-2, -0.9]$ & $w_x \in [-0.99, -0.9]$\\ \hline

$100\theta_{MC}$ & $    1.03958_{-    0.00039-    0.00078}^{+    0.00039+    0.00079}$ & $    1.04042_{-    0.00033-    0.00068}^{+    0.00035 + 0.00068}$ & $    1.04051_{-    0.00037-    0.00084}^{+    0.00041+    0.00076}$ & $    1.04084_{-    0.00027-    0.00062}^{+    0.00032+    0.00060}$\\

$\Omega_b h^2$ & $    0.02206_{-    0.00013-    0.00026}^{+    0.00013+    0.00027}$ & $    0.02228_{-    0.00015-    0.00029}^{+    0.00014+    0.00029}$ & $    0.02230_{-    0.00016-    0.00030}^{+    0.00015+    0.00030}$ & $    0.02239_{-    0.00015-    0.00027}^{+    0.00014+    0.00027}$ \\

$\Omega_c h^2$ & $    0.1322_{-    0.0071-    0.0109}^{+    0.0049+    0.0111}$ & $    0.1217_{-    0.0037-    0.0052}^{+    0.0024+    0.0059}$ & $    0.1202_{-    0.0040-    0.0064}^{+    0.0024+    0.0103}$ & $    0.1167_{-    0.0012-    0.0023}^{+    0.0012+    0.0023}$\\

$\tau$ & $    0.0429_{-    0.0175-    0.0317}^{+    0.0162+    0.0292}$ & $0.0646_{-    0.0163-    0.0345}^{+    0.0162+    0.0329}$ & $    0.0665_{-    0.0186-    0.0362}^{+    0.0184+    0.0353}$ & $    0.0776_{-    0.0169-    0.0345}^{+    0.0168+    0.0326}$\\

$n_s$ & $    0.9681_{-    0.0037-    0.0071}^{+    0.0036+    0.0075}$ & $ 0.9746_{-    0.0039-    0.0087}^{+    0.0046+    0.0078}$ & $    0.9763_{-    0.0043-    0.0083}^{+    0.0044+    0.0086}$ & $    0.9790_{-    0.0043-    0.0071}^{+    0.0041+    0.0076}$\\

${\rm{ln}}(10^{10} A_s)$ & $    3.0322_{-    0.0335-    0.0636}^{+    0.0319+    0.0622}$ & $    3.0702_{-    0.0323-    0.0655}^{+    0.0322+    0.0633}$ & $    3.0726_{-    0.0362-    0.0719}^{+    0.0358+    0.0685}$ & $    3.0922_{-    0.0328-    0.0702}^{+    0.0367+    0.0676}$ \\

$e_{\pi}$ & $   -0.0361_{-    0.0269-    0.0587}^{+    0.0312+    0.0531}$ & $   -0.0080_{-    0.0192-    0.0418}^{+    0.0196+    0.0391}$ & $   -0.0062_{-    0.0207-    0.0392}^{+    0.0202+    0.0426}$ & $    0.0057_{-    0.0271-    0.0410}^{+    0.0179+    0.0462}$\\

$w_x$ & $   -1.2099_{-    0.0018-    0.0196}^{+    0.0099+    0.0099}$ & $   -1.0601_{-    0.0262-    0.0654}^{+    0.0431+    0.0601}$ & $   -1.0391_{-    0.0402-    0.1084}^{+    0.0399+    0.0862}$ & $   -0.9749_{-    0.0150-    0.0151}^{+    0.0045+    0.0193}$\\

$\xi$ & $    0.1729_{-    0.1062-    0.1729}^{+    0.0948+    0.1574}$ & $    0.1574_{-    0.1574-    0.1574}^{+    0.0416+    0.2518}$ & $    0.1543_{-    0.1160-    0.1543}^{+    0.0740+    0.1717}$ & $    0.0401_{-    0.0401-    0.0401}^{+    0.0103+    0.0480}$\\

$\Omega_{m0}$ & $    0.3048_{-    0.0182-    0.0298}^{+    0.0133+    0.0313}$ & $    0.3037_{-    0.0079-    0.0147}^{+    0.0077+    0.0152}$ & $    0.3036_{-    0.0069-    0.0138}^{+    0.0070+    0.0141}$ & $    0.3065_{-    0.0058-    0.0122}^{+    0.0059+    0.0125}$ \\

$\sigma_8$ & $    0.7811_{-    0.0355-    0.0672}^{+    0.0376+    0.0686}$ & $    0.8074_{-    0.0182-    0.0352}^{+    0.0181+    0.0336}$ & $    0.8079_{-    0.0140-    0.0393}^{+    0.0213+    0.0341}$ & $    0.8125_{-    0.0128-    0.0259}^{+    0.0147+    0.0245}$ \\

$H_0$ & $   71.3016_{-    0.6416-    1.2947}^{+    0.6542+    1.2629}$ & $   69.0066_{-    0.9014-    1.4568}^{+    0.6404+    1.6489}$ & $   68.6800_{-    0.8911-    1.7090}^{+    0.8318+    1.8753}$ & $   67.5236_{-    0.4792-    1.0026}^{+    0.4626+    0.9725}$\\

\hline\hline
\end{tabular}
\label{tab:different-wx}
\end{table*}
\end{center}
\endgroup

\begin{figure*}
\includegraphics[width=0.35\textwidth]{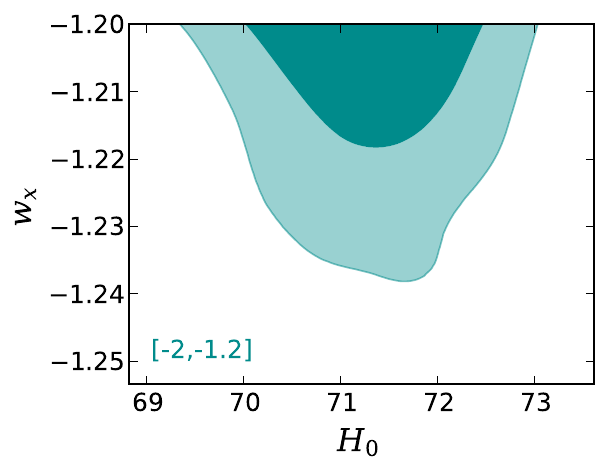}
\includegraphics[width=0.35\textwidth]{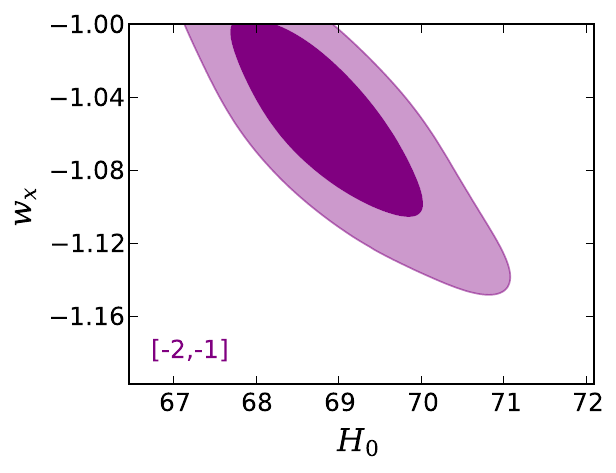}
\includegraphics[width=0.35\textwidth]{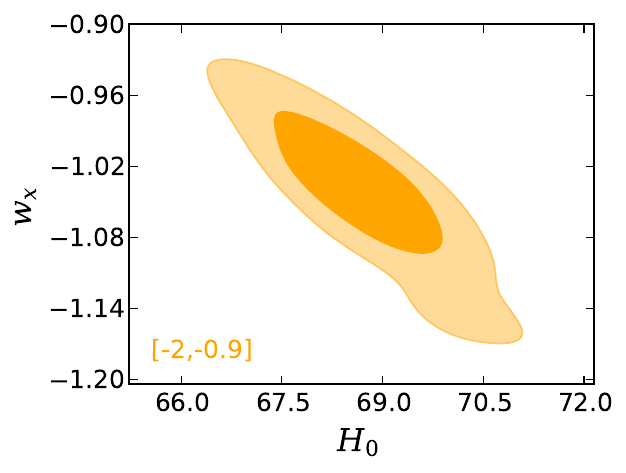}
\includegraphics[width=0.34\textwidth]{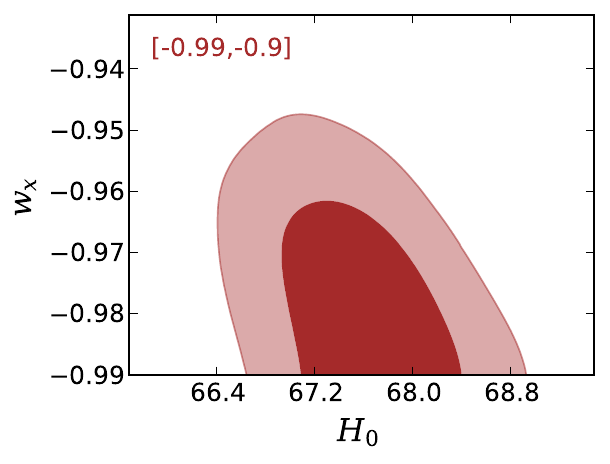}
\caption{68\% and 95\% confidence level contour plots in the fixed plane $(H_0, w_x)$ for different regions of the dark energy state parameter $w_x$. The combined analysis for this analysis has been fixed to be CMB+BAO+RSD+WL+HST+JLA+CC. }
\label{fig-H0}
\end{figure*}

\subsection{Easing the tension on $H_0$?}
\label{subsection-tension}

We now investigate whether the tension
on $H_0$ is released in this context. The tension
is one of the most talkative issues at current cosmological research.
However, at the very beginning, we recall what exactly the tension on $H_0$ is.
From the estimated values of $H_0$, one from Planck \cite{Ade:2015xua} (assuming $\Lambda$CDM as the base model) and one from Riess et al. \cite{Riess:2016jrr} (using the data from Hubble Space Telescope) one can see that
the values conflict amongst each other with a sufficient difference between their measurements. From Planck, the estimation of the current Hubble constant is $H_0= 67.27 \pm 0.66$ km~s$^{-1}$~Mpc$^{-1}$, while the same is reported in
\cite{Riess:2016jrr} having $H_0 = 73.24 \pm 1.74$ km~s$^{-1}$~Mpc$^{-1}$. That means the $H_0$ from \cite{Riess:2016jrr} is about $3\sigma$ higher from Planck's estimation. This is usually known as the tension on $H_0$. In the context of interacting dark energy, some latest articles  \cite{Kumar:2017dnp, DiValentino:2017iww} argued that the allowance of such coupling in the dark sectors becomes efficient to release such tension. Indeed this is a very potential result because the allowance of extra degrees of freedom in terms of the coupling strength might be able to release such tension. The difference of the earlier works \cite{Kumar:2017dnp, DiValentino:2017iww} with the current one is very clear $-$ here we consider the anisotropic stress into the picture, thus, perhaps one may expect slightly different result, and this is the main motivation of this section.  Thus, in the first row of Table \ref{tab:tension}, we have shown the constraints on $H_0$ for the $w_x$CDM$+\xi + e_{\pi}$ scenario while the second row of the table shows the constraints on $w_x$CDM$+\xi$ scenario (i.e. here $e_{\pi} =0$, that means no anisotropic stress). One can clearly see that the error bars on $H_0$ for the CMB analysis are extremely high in compared to the other analyses performed. It apparently releases the tension on the $H_0$ within the 68\% confidence-level. On the other hand, from the other analyses, presented in Table \ref{tab:tension}, one can clearly see that the combination CMB+BAO+WL+HST can only relieve the tension on $H_0$ within 68\% confidence-level while the other combinations do not look promising even at 99\% confidence-level. This phenomenon is true for both the interacting scenarios, that means interacting scenario with and without the anisotropic stress.

We investigate this issue more crucially taking the following approach. We consider the interacting scenario $w_x$CDM$+\xi+e_{\pi}$, for different regions of the dark energy state parameter $w_x$ setting its prior to the following four regions, namely, (i) $w_x \in [-2, -1.2]$, (ii) $w_x \in [-2, -1]$, (iii) $w_x \in [-2, -0.9]$ and finally (iv) $w_x \in [-0.99, -0.9]$. The analyses for the above four choices have been shown in Table \ref{tab:different-wx}. One can clearly conclude that as long as the dark energy equation of state remains in the phantom region (i.e. for $w_x \in [-2, -1.2]$), the tension on $H_0$ can be easliy released that fully supports a recent analysis \cite{DiValentino:2017iww} finding the same conclusion. While for the other regions of $w_x$, we do not find any significant signal for the alleviation of the tension on $H_0$. We have also shown this result in Fig. \ref{fig-H0}.

\subsection{Bayesian Evidence}
\label{sec-bayesian-evid}

We close the observational analysis of this work with a comparison of the interacting 
dark energy scenarios through the Bayesian evidence, an effective approach that enables us to judge the acceptance of the cosmological models compared to some reference model. As usual we adopt the reference model to be the six-parameters based $\Lambda$CDM cosmological model. To calculate the Bayesian evidence we need the posterior probability of the model parameters denoted by $\Theta$, given a specific astronomical data set $x$ for analyzing the model, any prior information and the model $M$. Using the Bayes theorem
\begin{eqnarray}\label{BE}
p(\Theta|x, M) = \frac{p(x|\Theta, M)\,\pi(\Theta|M)}{p(x|M)}
\end{eqnarray}
where $p(x|\Theta, M)$ is the likelihood function (this depends on the model parameters $\Theta$ with the fixed astronomical data set having $\pi(\Theta|M)$ as the prior information).  The quantity $p(x|M)$ located in the denominator of the right hand side of eqn. (\ref{BE}) is used for the model comparison and actually, this is the integral 
over the unnormalised posterior $\tilde{p} (\Theta|x, M) \equiv p(x|\Theta,M)\,\pi(\Theta|M)$ as follows: 
$E \equiv p(x|M) = \int d\Theta\, p(x|\Theta,M)\,\pi(\Theta|M)$, which is
also cited as the marginal likelihood. Now, let us consider any two models $M_i$ and $M_j$, where $M_i$ is the model that we want to compare with the reference model $M_j$ 
(this is the $\Lambda$CDM model under consideration). For this case, the posterior probability is given by

\begin{eqnarray}
\frac{p(M_i|x)}{p(M_j|x)} = \frac{\pi(M_i)}{\pi(M_j)}\,\frac{p(x| M_i)}{p(x|M_j)} = \frac{\pi(M_i)}{\pi(M_j)}\, B_{ij}.
\end{eqnarray}
where $B_{ij} = \frac{p(x| M_i)}{p(x|M_j)}$, is the Bayes factor of the model $M_i$ compared to $M_j$, the reference model.  This factor characterizes the observational viability of the model under considerationas follows: For $B_{ij} > 1 $, the astronomical data favor $M_i$ more strongly than $M_j$. For distinct measures of $B_{ij}$, sometimes we use $\ln B_{ij}$ for quantification, we follow the Jeffreys scales \cite{Kass:1995loi} summarized in Table \ref{tab:jeffreys}. 
The Bayesian evidence is calculated 
directly from the MCMC chains, the chains used to extract the parameters space of the cosmological models. A detailed description to calculate the Bayesian evidence of any cosmological model can be found at \cite{Heavens:2017hkr,Heavens:2017afc}. We use the publicly available code \texttt{MCEvidence}\footnote{Anyone can freely access the code from \href{https://github.com/yabebalFantaye/MCEvidence}{github.com/yabebalFantaye/MCEvidence}.}

In Table \ref{tab:bayesian}, we summarize the calculated values of $\ln B_{ij}$ for the interacting scenarios (with and without the anisotropic stress) compared to the reference model $\Lambda$CDM. The negative values in $\ln B_{ij}$  designate that the $\Lambda$CDM model is favored over the interacting scenarios.  We find that for almost all the observational data, the $\Lambda$CDM is strongly favored over both the interacting scenarios (with and without the anisotropic stress).

\begingroup                                                                                                                     
\squeezetable                                                                                                                   
\begin{center}                                                                                                                  
\begin{table}
\begin{tabular}{cc}                                                                                                            
\hline\hline                                                                                                                    
$\ln B_{ij}$ & Strength of evidence for model ${M}_i$ \\ \hline
$0 \leq \ln B_{ij} < 1$ & Weak \\
$1 \leq \ln B_{ij} < 3$ & Definite/Positive \\
$3 \leq \ln B_{ij} < 5$ & Strong \\
$\ln B_{ij} \geq 5$ & Very strong \\
\hline\hline                                                                                                                    
\end{tabular}                                                                                                                   
\caption{Revised Jeffreys scale quantifying the observational viability of the model $M_i$ compared to the reference model $M_j$.} \label{tab:jeffreys}                                                                                                   
\end{table}                                                                                                                     
\end{center}                                                                                                                    
\endgroup

\begingroup                                                                                                                     
\begin{center}                                                                                                                  
\begin{table*}
\begin{tabular}{cccccc}                                                                                                            
\hline\hline                                                                                                                    
Dataset & Model & $\ln B_{ij}$ & Strength of evidence for the $\Lambda$CDM model \\ \hline
CMB  & IDE with anisotropic stress &  $-3.9$ & Strong \\
CBR   & IDE with anisotropic stress &  $-2.2$ & Definite/Positive \\
CBRH   & IDE with anisotropic stress &  $-3.8$ & Strong \\
CBWH   & IDE with anisotropic stress &  $-5.2$ & Very Strong\\
CBRWHJC  & IDE with anisotropic stress &  $-4.9$ & Strong \\
\hline

CMB & IDE with no anisotropic stress  &  $-3.1$ & Strong \\
CBR & IDE with no anisotropic stress  &   $-1.9$ &  Definite/Positive \\
CBRH & IDE with no anisotropic stress  &  $-3.6$ & Strong \\
CBWH & IDE with no anisotropic stress  &   $-5.3$ & Very Strong \\
CBRWHJC & IDE with no anisotropic stress  & $-4.0$ &  Strong \\

\hline\hline  

\end{tabular}                                                                                                                   
\caption{The table summarizes the 
values of $\ln B_{ij}$ for the interacting scenarios (with and without the anisotropic stress) compared to the reference model $\Lambda$CDM model for different data sets. 
The negative values of $\ln B_{ij}$ according to the Bayesian point of view indicate the preference of $\Lambda$CDM model over the interacting scenarios. Here,  
CBR = CMB+BAO+RSD, CBRH = CMB+BAO+RSD+HST, CBWH = CMB+BAO+WL+HST, CBRWHJC = CMB+BAO+RSD+WL+HST+JLA+CC. }\label{tab:bayesian}                                                                                                   
\end{table*}                                                                                                                     
\end{center}                                                                                                                    
\endgroup

\section{Summary and Conclusions}
\label{sec-discuss}

For the first time, we consider an interaction scenario between pressureless dark matter and dark energy when a matter-sourced anisotropic stress is present into the formalism. In general, the contribution from the anisotropic stress is often excluded from the cosmic picture, but however, a complete cosmological scenario must include all the associated parameters where theoretically, there is no such strong reason to exclude such anisotropic stress. And from the observational point of view, only the analyses might tell us whether the inclusion of anisotropic stress is necessary or not. Thus, keeping the anisotropic stress into our discussions, we try to explore this general interacting scenario.
The dark energy equation of state, $w_x$, in this work has been considered to be time independent, and the interaction rate, $Q$, has been taken to be of the form $Q = 3 H \xi (1+w_x) \rho_x$ \cite{Yang:2017zjs}
in order to investigate the entire parameter space for $w_x$ unlike other interaction models where two separate regions for the dark energy equation of state are considered, see \cite{Yang:2017zjs, Yang:2017ccc} for a detailed motivation behind the choice of  the above interaction rate. The cosmological scenario has been constrained for different combinations of the astronomical data with latest compilation (see Table \ref{tab:results}).

Our analyses show that the current observational data indicate for a {\it nonzero interaction in the dark sectors} with a {\it nonzero anisotropic stress} in addition. That means, from the observational base, $e_{\pi}$ should not be identically taken to be zero to explore the dynamical features of the universe.
Interestingly, most of the combined analyses include $\xi= 0$ and $e_{\pi} = 0$ within the 68\% confidence-region which means that at the background level, the model could mimick the non-interacting $w_x$CDM model. And moreover, from the estimated values of $w_x$ from different combined analyses, one can also see that $w_x$ is very close to `$-1$' boundary meaning that the model is actually close to the $\Lambda$CDM cosmological model as well. 
However, the most striking result is observed from the perturbative analysis which reports that {\it the model is different from the $\Lambda$CDM model}. 
We find that if we allow $\xi$ to be very small (even if we assume $\xi=0$) but consider the anisotropic stress whatever small its strength be, a deviation from $\Lambda$-cosmology is pronounced from the ratio of CMB TT spectra (see the right panel of Fig. \ref{fig:cmbTT1}). On the other hand, even if we make $e_{\pi} = 0$ and consider different strengths of the interactions (see Fig. \ref{fig:cmbTT2}), then the interaction model shows a deviation from the $\Lambda$-cosmology that is only perfectly realized from the ratio of the CMB TT spectra displayed in the right panel of Fig. \ref{fig:cmbTT2}. {\it This is an interesting result because  from the background analysis we could not distinguish} the interaction model from the base $\Lambda$CDM while only the analysis at the perturbative level became able to find out such differences. 

We also find that the tension on $H_0$ can be alleviated. Actually, whenever interaction
is present, then the release of tension on $H_0$ is possible as found in some latest
investigations \cite{Kumar:2016zpg, DiValentino:2017iww} where the authors show that the coupling into the dark sector shifts the Hubble parameter value toward its local measurement. Since in the current work we consider the anisotropic stress into the picture, hence, we have investigated how the presence of an anisotropic stress controls the tension on $H_0$. The values of $H_0$ from different analyses have been shown in Table \ref{tab:tension} which clearly shows that the combined analysis CMB+BAO+WL+HST could alleviate the tension on the $H_0$. The analysis with only CMB allow a very large region of $H_0$ even at 68\% confidence-level and thus naturally, the tension is found to be released. While for the other combinations, we do not observe anything similar to that. But, interestingly enough, we find that if we allow $w_x$ to lie within the phantom region only, the tension is surely released (see the second column of Table \ref{tab:different-wx}). This result coincides with a latest investigation \cite{ DiValentino:2017iww}, although the major difference with this work is that, here we have an extra degrees of freedom in terms of the anisotropic stress.  However, from the Bayesian analysis, we find that the $\Lambda$CDM model is well favored over the present interacting scenarios. 

As a closing remark, a number of investigations might be performed following the present work. In particular, it is interesting to see the behavior of the interacting scenario in presence of a dynamical $w_x$ instead of its constant value. The inclusion of massive neutrinos is another important addition in this picture. As the consideration of anisotropic stress is new in the context of coupled dark matter $-$ dark energy models, one can explore some more interesting and important ideas. We hope to address some of them in near future, although such investigations are open to all.

\section*{ACKNOWLEDGMENTS}
It is a pleasure to thank the referee of MNRAS for several comments to improve the work. 
The authors acknowledge the use of publicly available markov
chain monte carlo package \texttt{cosmomc}.  
WY acknowledges the support from the National Natural Science Foundation of China under Grants No. 11705079 and No. 11647153.
LX acknowledges the support from the National Natural Science Foundation of China under Grants No. 11275035, No.11675032, and ``the Fundamental Research Funds for the Central Universities'' under Grant No. DUT16LK31.
DFM acknowledges the support from the Research Council of Norway, and this paper is based upon work from COST action CA15117 (CANTATA), supported by COST (European Cooperation in Science and Technology).

\end{document}